\documentclass{aastex61}
\usepackage{CJKutf8}

\newcommand{\nt}{34~}

\newcommand{\ntno}{838~}

\submitjournal{AAS}

\shorttitle{OSSOS Twotino}
\shortauthors{Chen et al.}

\newcommand{\phit}{\phi_{21}}
\newcommand{\dphit}{\Delta\phi_{21}}

\begin{document}

\title{OSSOS XVIII: Constraining migration models with the 2:1 resonance using the Outer Solar System Origins Survey}


\correspondingauthor{Ying-Tung Chen}
\email{ytchen@asiaa.sinica.edu.tw}

\author[0000-0001-7244-6069]{Ying-Tung Chen (\begin{CJK*}{UTF8}{bkai}陳英同\end{CJK*})}
\affiliation{Institute of Astronomy and Astrophysics, Academia Sinica, Taipei 10617, Taiwan}

\author{Brett Gladman}
\affiliation{Department of Physics and Astronomy, University of British Columbia, 6224 Agricultural Road, Vancouver, BC V6T 1Z1, Canada}

\author[0000-0001-8736-236X]{Kathryn Volk}
\affiliation{Lunar and Planetary Laboratory, The University of Arizona, 1629 E University Blvd, Tucson, AZ 85721, USA}

\author[0000-0001-5061-0462]{Ruth Murray-Clay}
\affiliation{Department of Astronomy and Astrophysics, University of California, Santa Cruz, CA 95064, USA}

\author[0000-0003-4077-0985]{Matthew J. Lehner}
\affiliation{Institute of Astronomy and Astrophysics, Academia Sinica, Taipei 10617, Taiwan}
\affiliation{Department of Physics and Astronomy, University of Pennsylvania, 209 S. 33rd St., Philadelphia, PA 19125,
USA}
\affiliation{Harvard-Smithsonian Center for Astrophysics, 60 Garden St., Cambridge, MA 02138, USA}

\author[0000-0001-7032-5255]{J. J. Kavelaars}
\affiliation{Herzberg Astronomy and Astrophysics Research Centre, National Research Council of Canada, 5071 West Saanich Rd, Victoria, British Columbia V9E 2E7, Canada}
\affiliation{Department of Physics and Astronomy, University of Victoria, Elliott Building, 3800 Finnerty Rd, Victoria, BC V8P 5C2, Canada}

\author{Shiang-Yu Wang (\begin{CJK*}{UTF8}{bkai}王祥宇\end{CJK*})}
\affiliation{Institute of Astronomy and Astrophysics, Academia Sinica, Taipei 10617, Taiwan}

\author[0000-0001-7737-6784]{Hsing-Wen Lin (\begin{CJK*}{UTF8}{bkai}林省文\end{CJK*})} 
\affiliation{Department of Physics, University of Michigan, Ann Arbor, MI 48109, USA}
\affiliation{Institute of Astronomy, National Central University, 32001, Taiwan}

\author[0000-0003- 0926-2448]{Patryk Sofia Lykawka}
\affiliation{School of Interdisciplinary Social and Human Sciences, Kindai University, Japan}

\author[0000-0003-4143-8589]{Mike Alexandersen} 
\affiliation{Institute of Astronomy and Astrophysics, Academia Sinica, Taipei 10617, Taiwan}

\author[0000-0003-3257-4490]{Michele T. Bannister}
\affiliation{Astrophysics Research Centre, School of Mathematics and Physics, Queen's University Belfast, Belfast BT7 1NN, United Kingdom}

\author[0000-0001-5368-386X]{Samantha M. Lawler}
\affiliation{Herzberg Astronomy and Astrophysics Research Centre, National Research Council of Canada, 5071 West Saanich Rd, Victoria, British Columbia V9E 2E7, Canada}

\author[0000-0001-9677-1296]{Rebekah I. Dawson}
\affiliation{Department of Astronomy \& Astrophysics, Center for Exoplanets and Habitable Worlds, The Pennsylvania State University, University Park, PA 16802, USA}

\author{Sarah Greenstreet}
\affiliation{B612 Asteroid Institute, 20 Sunnyside Ave, Suite 427, Mill Valley, CA 94941}
\affiliation{DIRAC Center, Department of Astronomy, University of Washington, 3910 15th Ave NE, Seattle, WA 98195}

\author{Stephen D. J. Gwyn}
\affiliation{Herzberg Astronomy and Astrophysics Research Centre, National Research Council of Canada, 5071 West Saanich Rd, Victoria, British Columbia V9E 2E7, Canada}

\author[0000-0003-0407-2266]{Jean-Marc Petit}
\affiliation{Institut UTINAM UMR6213, CNRS, Univ. Bourgogne Franche-Comt\'e, OSU Theta F25000 Besan\c{c}on, France}

\begin{abstract}
Resonant dynamics plays a significant role in the past evolution and current state of our outer Solar System.
The population ratios and spatial distribution of Neptune's resonant populations are direct clues to understanding the history of our planetary system.
The orbital structure of the objects in Neptune's 2:1  mean-motion resonance (\emph{twotinos}) has the potential to be a tracer of planetary migration processes.
Different migration processes produce distinct architectures, recognizable by well-characterized surveys.
However, previous characterized surveys only discovered a few twotinos, making it impossible to model the intrinsic twotino population.

With a well-designed cadence and nearly 100\% tracking success, the Outer Solar System Origins Survey (OSSOS) discovered 838 trans-Neptunian objects, of which 34 are securely twotinos with well-constrained libration angles and amplitudes.
We use the OSSOS twotinos and the survey characterization parameters via the OSSOS Survey Simulator to inspect the intrinsic population and orbital distributions of twotino. 
The estimated twotino population, 4400$^{+1500}_{-1100}$ with $H_r<8.66$ (diameter$\sim$100km) at 95\% confidence, is consistent with the previous low-precision estimate. 
We also constrain the width of the inclination distribution to a relatively narrow value of $\sigma_i$=$6\arcdeg^{+1}_{-1}$, and find the eccentricity distribution is consistent with a Gaussian centered on $e_\mathrm{c}=0.275$ with a width $e_\mathrm{w}=0.06$. We find a single-slope exponential luminosity function with $\alpha=0.6$ for the twotinos.
Finally, we for the first time meaningfully constrain the fraction of symmetric twotinos, and the ratio of the leading asymmetric islands; both fractions are in a range of 0.2--0.6. These measurements rule out certain theoretical models of Neptune's migration history.
\end{abstract}

\keywords{celestial mechanics --- Kuiper belt: general --- surveys}

\section{Introduction} \label{sec:intro}
The small celestial bodies currently in the trans-Neptunian region are believed to be the remaining planetesimals that formed early in the history of the Solar System.
Based on these surviving bodies, many theoretical models of the outer Solar System's early dynamical history, including the migration of Neptune to its current orbit, have been constructed in attempts to reproduce the current structure and populations of trans-Neptunian objects (TNOs) in the outer Solar System \citep[e.g.,][]{mal95, lev03, bra13, nes16}.
Different modes for planetary migration can produce different architectures in the present outer Solar System. 
In particular, differences in the pre-migration structure of the planetesimal disk, the speed and smoothness of Neptune's migration, and the dynamical excitation of Neptune's orbit during migration can all leave signatures in the current TNO population \citep[e.g.,][]{hah05, mur05, lev08, nes15a, nes15b, nes16}. 

During planetary migration, two of Neptune's strongest resonances, the 3:2 resonance (whose members are referred to as \emph{plutinos}) and the 2:1 resonance (whose members are referred to as \emph{twotinos}), would have captured a large number of planetesimals; these two resonances contain large numbers of observed resonant objects.
The 2:1 resonance with Neptune, which is roughly located at semimajor axis $a$ = 47.8~au, lies near the outer edge of the main classical belt \citep{gla08}.
The spatial and orbital distribution of the current twotino population provides clues for understanding the pre-migration planetesimal disk and the history of Neptune's migration. 
In addition to the usual eccentricity and inclination distributions, the 2:1 resonance's multiple libration islands provide an additional constraint on the Solar System's dynamical history. 
Like other $n$:1 mean motion resonances, the 2:1 resonance has three stable libration centers \citep[see, e.g.,][]{bea94,mal96}: the `symmetric' libration center, where objects' perihelion location librates around a point centered $180^\circ$ away from Neptune, and two `asymmetric' libration centers, where objects' perihelion locations librate around points that lead or trail Neptune by $\sim60-100^\circ$.
In simulations of resonant capture during planetary migration, the orbital distribution and relative population ratios of the two asymmetric islands (the leading and trailing islands) within the 2:1 can differ depending on migration parameters \citep{chi02, mur05}.
In particular, the distribution of the twotino population can provide an important constraint on the speed and total distance of Neptune's migration if this population is sufficiently constrained by outer Solar System surveys.

Past observational surveys have suggested that the twotino population does differ between the libration islands, but only with rough estimates.
The Deep Ecliptic Survey (DES) was a pioneering project to systematically survey the ecliptic plane in the early 2000s.
This project discovered five twotinos \citep{ell05}.
Combining the DES discoveries and other known twotinos, the observationally biased results showed that two twotinos lie at longitudes behind that of Neptune, and seven are located ahead of it, but interpretation of the asymmetry is difficult due to incomplete knowledge of survey bias.
A rough estimate of the simple debiased trailing-to-leading ratio (3:6) excludes a very rapid migration (migration timescale $\tau \leq 10^{6}$~yr) history for Neptune \citep{mur05}.
A subsequent characterized\footnote{A `characterized' survey is one which provide a measured detection efficiency versus magnitude and rate for all its survey fields, along with a fully tracked sample (or one which the tracking efficiency can be modeled).} survey for outer Solar System objects was the Canada-France Ecliptic Plane Survey (CFEPS) \citep{kav09, pet11}. 
Combining the discovery of five twotinos and the survey characterization, CFEPS was only able to make a rough estimate of the twotino population and a simple comparison with DES results \citep{gla12}.
It is impractical to estimate the fraction of twotinos in the symmetric island and also the ratio of the leading and trailing islands with such a small number of detections.
Hence, wider and deeper surveys that detect more twotinos were necessary for understanding the intrinsic population and structure.

The Outer Solar System Origins Survey (OSSOS) was a Large Program with the 3.6~m Canada-France-Hawaii Telescope (CFHT), operating 2013--2017, which searched for TNOs in 155~deg.$^2$ of sky on and near the ecliptic plane \citep{ban16, ban18}.
The detailed characterization of OSSOS provides the capability to accurately constrain the intrinsic distributions of specific TNO sub-populations \citep{law18}.
\citet{vol16} used the twotino detections in the first quarter of OSSOS and in CFEPS (total: 4+5) by utilizing the OSSOS Survey Simulator to place constraints on the intrinsic number of twotinos and their distribution in the three libration islands. 
However, these constraints were relatively weak due to the small number of observed twotinos.
The full sample of \nt~OSSOS twotinos provide more precise constraints, which are extremely valuable for testing outer Solar System evolution models.
These observed twotinos have very precisely determined orbital parameters, including accurate resonant libration centers and amplitudes, due to the well-designed cadence of the OSSOS observations, and these parameters are key to estimating the resonance's intrinsic population and distribution for comparison to the end-states of theoretical models.

In this study, we extend the work of \citet{vol16} to the full OSSOS twotino sample, and we build an empirical model of the intrinsic twotino population based on the observed OSSOS twotinos and models of the twotino population from theoretical simulations.
This model is fine-tuned to match the orbital distributions of OSSOS twotinos using the Survey Simulator.
We are thus able to use the OSSOS detections to provide first quantitative constraints on the twotino orbital distribution, including the fraction of the population belonging to each libration island.

Section~\ref{sec:obs} of this paper describes the observations and characterization of the OSSOS twotino sample. 
Section~\ref{sec:model} describes the basic dynamics of the twotino population and details the empirical model.
In Section~\ref{sec:verify}, we examine how well this empirical model matches the OSSOS observations using a statistical analysis of the population and constrain the fraction of the twotino population in each of the libration islands. 
In Section~\ref{sec:dis}, we discuss the connections between our observations and theoretical models.
A summary of this study is provided in Section~\ref{sec:sum}.

\begin{figure}[!ht]
\centering
\plotone{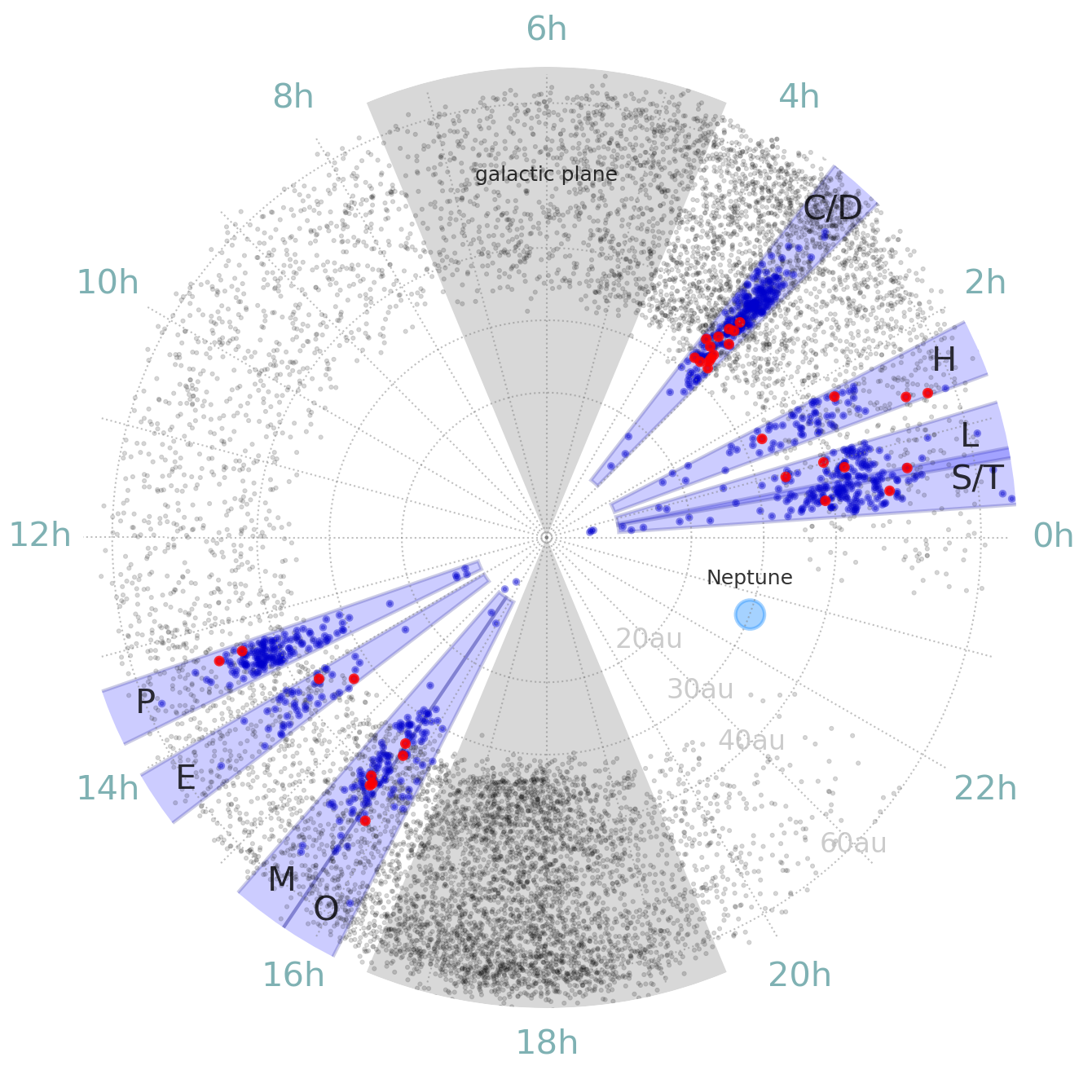}
\caption{The spatial distribution of the OSSOS observational blocks (blue sectors) and detected TNOs (blue dots) are shown projected onto the ecliptic plane. The detected twotinos are shown as red dots projected onto the ecliptic plane, and the gray dots are the empirical model of the intrinsic twotino population in this study (see Section~\ref{sec:model}). The gray shaded areas indicate the location of the Galactic plane along the ecliptic where TNO detection is very difficult due to the high background star density. Neptune's indicated location is at the rough mid-point of the full survey (2014/07/01). Note that the upper right quadrant is (northern) ``fall''.}
\label{fig:ossos_top_view}
\end{figure}

\section{Observations} \label{sec:obs}

\begin{deluxetable*}{cc|ccccccccl}
\tablecaption{OSSOS TNOs in Neptune's 2:1 Resonance\tablenotemark{a} \label{tab:tab21}}
\tablecolumns{11}
\tablehead{
\multicolumn{2}{c}{Designations} &
\colhead{$a$} &
\colhead{$e$} &
\colhead{$i$} &
\colhead{$d$\tablenotemark{b}} &
\colhead{$\langle\phit\rangle$} &
\colhead{$A_{\phit}$} &
\colhead{Mag} &
\colhead{$H$} &
\colhead{Libration Island}
\\
\cline{1-2}
\colhead{OSSOS} &
\colhead{MPC} &
\colhead{(au)} &
\colhead{} &
\colhead{($\arcdeg$)} &
\colhead{(au)} &
\colhead{($\arcdeg$)} &
\colhead{($\arcdeg$)} &
\colhead{($r$)} &
\colhead{($r$)} &
\colhead{}
}
\startdata
o5c025 & 2015 VA$_{166}$ & 47.761 & 0.41828 & 13.153 & 35.2 & 65.78 & 8$^{+3}_{-3}$ & 24.30 &  8.8 & leading \\
o5c016 & 2015 VZ$_{165}$ & 47.689 & 0.36811 & 4.115 & 32.3 & 67.48 & 8$^{+2}_{-2}$ & 22.97 &  7.9 & leading \\
o5d008 & 2015 VD$_{166}$ & 47.671 & 0.35856 & 1.911 & 32.2 & 68.29 & 11.0$^{+0.9}_{-0.7}$ & 24.37 &  9.3 & leading \\
o5d014 & 2015 VH$_{166}$ & 47.750 & 0.31534 & 3.390 & 33.6 & 70.39 & 3.9$^{+0.7}_{-0.8}$ & 24.09 &  8.8 & leading \\
o5d026 & 2015 VF$_{166}$ & 47.701 & 0.31562 & 1.569 & 36.5 & 70.89 & 8$^{+2}_{-3}$ & 24.39 &  8.8 & leading \\
o5d021 & 2015 VD$_{157}$ & 47.702 & 0.31024 & 0.892 & 34.7 & 71.15 & 7.6$^{+1.4}_{-1.6}$ & 24.34 &  8.9 & leading \\
o5c031 & 2015 VY$_{165}$ & 47.733 & 0.30280 & 1.288 & 36.7 & 71.28 & 5.5$^{+0.2}_{-0.1}$ & 23.20 &  7.5 & leading \\
o5d010 & 2015 VL$_{166}$ & 47.796 & 0.32197 & 5.457 & 33.1 & 71.29 & 12.1$^{+0.5}_{-1.0}$ & 24.93 &  9.7 & leading \\
o5c014 & 2015 VB$_{166}$ & 47.647 & 0.32663 & 1.045 & 32.2 & 72.62 & 19.5$^{+0.6}_{-2.2}$ & 23.61 &  8.5 & leading \\
o3l80 & 2013 SA$_{101}$ & 47.752 & 0.27033 & 3.689 & 50.7 & 77.32 & 20.3$^{+1.4}_{-0.3}$ & 23.82 &  6.8 & leading \\
o4h15 & 2014 UH$_{228}$ & 47.627 & 0.32585 & 3.110 & 32.7 & 77.34 & 30.3$^{+1.6}_{-0.6}$ & 23.94 &  8.8 & leading \\
o5d019 & 2015 VJ$_{166}$ & 47.595 & 0.29551 & 3.653 & 34.2 & 78.17 & 28.6$^{+0.7}_{-0.9}$ & 24.40 &  9.1 & leading \\
o5d039 & 2015 VG$_{166}$ & 47.805 & 0.20982 & 17.175 & 38.5 & 78.27 & 18.6$^{+1.0}_{-0.8}$ & 24.86 &  9.0 & leading \\
o3l34 & 2013 SQ$_{100}$ & 47.847 & 0.25342 & 12.957 & 42.3 & 81.07 & 30.5$^{+1.6}_{-2.1}$ & 23.64 &  7.4 & leading \\
o5d035 & 2015 VE$_{166}$ & 47.836 & 0.21248 & 4.344 & 38.3 & 85.83 & 27.6$^{+1.8}_{-1.8}$ & 24.96 &  9.1 & leading \\
o4h43 & 2014 US$_{228}$ & 47.941 & 0.32533 & 7.165 & 44.3 & 90.8 & 47.6$^{+2.2}_{-6.0}$ & 24.07 &  7.6 & leading \\
o3l08 & 2013 UH$_{17}$ & 47.694 & 0.30232 & 9.332 & 34.1 & 91.43 & 46.6$^{+1.4}_{-3.6}$ & 24.35 &  8.9 & leading\tablenotemark{c}\\
\cline{1-11}
o3o33 & 2013 JJ$_{64}$ & 47.768 & 0.08252 & 7.650 & 46.5 & 175.36 & 113.2$^{+2.2}_{-18.3}$ & 23.99 &  7.3 & symmetric\tablenotemark{d} \\
o5m57 & 2015 KX$_{173}$ & 47.911 & 0.23826 & 15.529 & 41.6 & 179.72 & 148.9$^{+0.9}_{-1.5}$ & 24.58 &  8.4 & symmetric \\
o3e55 & 2013 GX$_{136}$ & 48.003 & 0.25156 & 1.100 & 37.0 & 179.87 & 155.0$^{+0.4}_{-0.6}$ & 23.41 &  7.7 & symmetric \\
o5d055 & 2015 VK$_{166}$ & 48.039 & 0.20252 & 2.323 & 39.9 & 179.87 & 160.2$^{+1.1}_{-1.1}$ & 23.20 &  7.2 & symmetric \\
o3l17 & 2013 UZ$_{17}$ & 47.532 & 0.17602 & 7.179 & 39.6 & 179.88 & 152.6$^{+0.6}_{-1.5}$ & 23.98 &  7.9 & symmetric \\
o5m60 & 2015 KL$_{167}$ & 47.440 & 0.25558 & 10.304 & 42.1 & 179.91 & 153.5$^{+2.0}_{-2.6}$ & 24.47 &  8.2 & symmetric \\
o5d027 & 2015 VC$_{166}$ & 48.105 & 0.24046 & 2.917 & 37.4 & 179.92 & 164.4$^{+11.1}_{-1.7}$ & 24.65 &  8.9 & symmetric \\
o5t47 & 2015 RX$_{277}$ & 47.642 & 0.20629 & 0.073 & 47.8 & 179.93 & 154.6$^{+1.4}_{-0.6}$ & 24.37 &  7.5 & symmetric \\
o5s27 & 2015 RY$_{277}$ & 47.457 & 0.27731 & 3.457 & 38.8 & 180.08 & 153.5$^{+3.3}_{-0.4}$ & 23.79 &  7.8 & symmetric \\
\cline{1-11}
o5m49 & 2015 KN$_{167}$ & 47.896 & 0.21359 & 6.029 & 40.9 & 257.05 & 50.7$^{+6.4}_{-5.1}$ & 24.34 &  8.2 & trailing \\
o5p121 & 2015 GY$_{51}$ & 47.549 & 0.30996 & 8.955 & 44.9 & 263.15 & 53.8$^{+46.0}_{-1.4}$ & 23.83 &  7.3 & trailing\tablenotemark{e} \\
o4h54 & 2014 UY$_{229}$ & 47.629 & 0.21375 & 3.403 & 56.3 & 275.89 & 27.4$^{+1.1}_{-1.0}$ & 24.63 &  7.1 & trailing \\
o4h53 & 2014 UW$_{229}$ & 47.725 & 0.17033 & 14.607 & 53.3 & 275.95 & 19.0$^{+0.5}_{-0.4}$ & 24.31 &  7.0 & trailing \\
o3e05 & 2013 GW$_{136}$ & 47.744 & 0.34410 & 6.660 & 33.0 & 277.25 & 39.1$^{+0.9}_{-0.3}$ & 22.69 &  7.4 & trailing \\
o5m26 & 2015 KM$_{167}$ & 47.763 & 0.33369 & 8.490 & 34.5 & 283.19 & 30.0$^{+1.4}_{-4.3}$ & 24.44 &  9.0 & trailing \\
o5p141 & 2015 GZ$_{54}$ & 47.756 & 0.27653 & 1.957 & 48.3 & 283.6 & 19.1$^{+1.2}_{-1.4}$ & 24.57 &  7.7 & trailing \\
o3o18 & 2013 JE$_{64}$ & 47.789 & 0.28524 & 8.335 & 36.1 & 284.12 & 22.5$^{+0.4}_{-0.5}$ & 23.56 &  7.9 & trailing \\
\enddata
\tablenotetext{a}{All digits given are significant, although $d$ and $H$ are more accurate than listed here.}
\tablenotetext{b}{Object distance at discovery.}
\tablenotetext{c}{Island switching after 5~Myr.}
\tablenotetext{d}{Island switching.}
\tablenotetext{e}{A few symmetric clones exist}

\end{deluxetable*}

\citet{ban16, ban18} provided a detailed description of the survey design and implementation for OSSOS.
The major characteristics of this survey are: (1) a total coverage area of  155~deg$^2$ over eight blocks on the sky (each $\approx$20~deg$^2$) within 10$\arcdeg$ of the ecliptic plane, (2) limiting magnitudes in the range 24.1--25.2 in $r$-band, (3) precovery and follow-up observations implemented with the same telescope before and after the discovery observations, and (4) precise characterization of the survey's detection efficiency.
The five year survey, with its fully characterized detection efficiency and complete tracking observations, obviates ephemeris bias and provides a unique approach to understanding intrinsic TNO populations.
In particular, one of the major science goals of OSSOS was to reveal the intrinsic resonant populations in the outer Solar System, so the survey block locations were chosen to optimize the detection of a large number of 3:2 and 2:1 objects, i.e. areas leading and trailing Neptune by 30--150$\arcdeg$ (see Figure~\ref{fig:ossos_top_view}).
In the full survey, OSSOS discovered \ntno~TNOs, including \nt~twotinos.
Each detected TNO was dynamically classified using numerical integrations according to the scheme outlined in \citet{gla08} and \citet{ban18}; the best-fit orbit is used for an object's nominal dynamical classification, and the minimum and maximum semimajor axis orbits consistent with the observations are used to determine whether that classification is secure.
Each OSSOS twotino was observed 24--47~times over total arc lengths of 1.1--4.1~years, which typically resulted in well-constrained orbits with only 0.1\% uncertainty at the most in semi-major axis; thus all of the OSSOS twotinos have secure classifications in the 2:1 resonance. 
The procedure to determine each twotino's libration center and libration amplitude (with uncertainty) is fully described in appendix B of \citet{vol16}; briefly, these parameters are determined by integrating 250 clones of each twotino that span the uncertainties associated with the best-fit orbit and analyzing their libration characteristics.
The orbital elements and resonance parameters of all OSSOS twotinos are listed in Table \ref{tab:tab21}.

\section{Twotino Model} \label{sec:model}

\subsection{Dynamics in the 2:1 resonance} \label{sec:dyna}
\begin{figure}[ht!]
\centering
\plotone{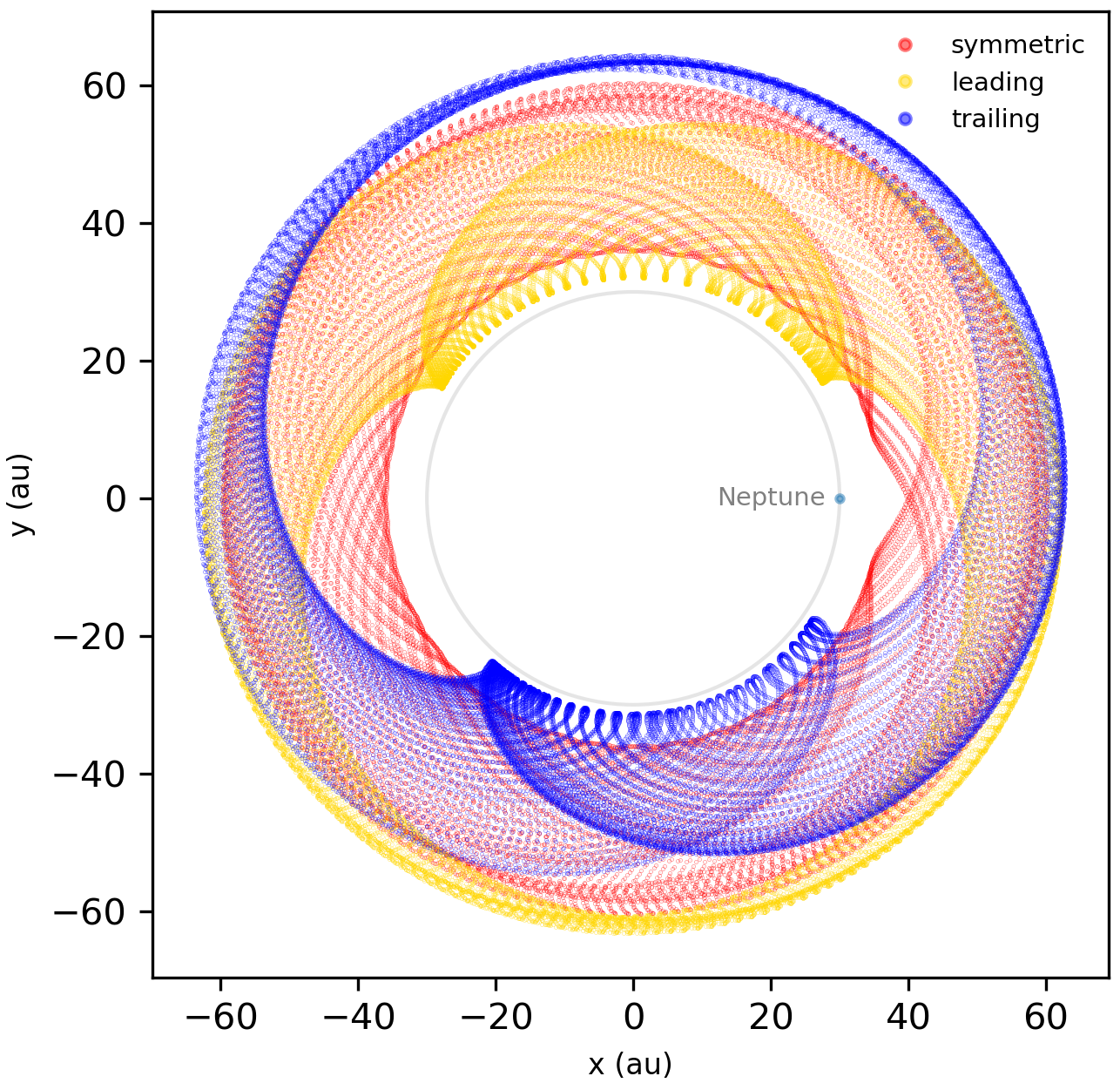}
\caption{An illustration of the 2:1 resonance's three libration islands using numerical simulations of three real OSSOS twotinos in a reference frame that co-rotates with Neptune (the single light blue dot). The red (o3e55/2013 GX136), yellow (o4h43/2014 US228) and blue (o3e05/2013 GW136) paths indicate libration in the symmetric, leading, and trailing islands, respectively. The step size between dots is about one year, and the total time span is about 29K years. 
The observational bias for detecting twotinos in the different libration islands as a function of longitude relative to Neptune is evident in this figure, as the objects are easier to detect near perihelion. }
\label{fig:three_islands}
\end{figure}

Occupancy of the 2:1 mean-motion resonance is diagnosed by a libration (oscillation) of the resonant argument:
\begin{equation}
\label{eq:resang}
\phit = 2 \lambda - \lambda_N - \varpi    \label{eq:resarg}
\end{equation} 
as time evolves.  Here the two mean longitudes ($\lambda$ and $\lambda_N$) are that of the TNO and Neptune, and $\varpi$ is the TNO's longitude of perihelion.  
We define the amplitude of the libration of this angle to be $\dphit = (\phit{_{max}} - \phit{_{min}})/2$.
For the 2:1 resonance there are three possible centers, or \emph{islands} (see Figure~\ref{fig:three_islands}), about which $\phit$ librates:

\begin{itemize}

\item \emph{Symmetric} libration around a central (roughly `average') $\langle\phit\rangle$=180$^\circ$, usually with an amplitude $\dphit $=140--170$^\circ$. While resonant libration is typically modeled as a sinusoidal process \citep[see, e.g.,][]{murray99}, high-amplitude libration such as that of the symmetric twotinos is more of a `triangular wave' (an example of this type of libration is shown in Appendix~C of \citealt{vol16}). This means all values of $\phit$ in the range $180 \pm \dphit$ are equally likely, so twotinos spend equal time at all allowed values of $\phit$.

\item \emph{Leading asymmetric} libration around a libration center $\langle\phit\rangle$ between $\sim60$ and 130$^\circ$; the exact libration center depends on the heliocentric orbital eccentricity $e$ \citep{nes01}.  The libration amplitude  around this central value depends on the location of the libration center and thus indirectly also on $e$.  

\item \emph{Trailing asymmetric} libration around a libration center $\langle\phit\rangle$ between $\sim230$ and 300$^\circ$, but otherwise identical to the leading asymmetric island.

\end{itemize}

The goal of this study is to model the distributions of these quantities in a way that agrees with the observations by OSSOS.
All surveys have detection biases (see \citet{gla12} for an detail discussion), but in the case of OSSOS we can analyse those biases and compare them with the measurements of the 2:1 population, showing that OSSOS has strong but well-understood biases in the detection of twotinos (see Figure~\ref{fig:three_islands}).
The comparison with observations requires subjecting a model of these distributions to the detection biases via a simulation of which types of objects are most likely to be detected by OSSOS.
So comparing a model of the intrinsic twotino distribution with observations requires exposing the model to the same biases as the observations.
We employ the OSSOS survey simulator (version 11) to simulate observational biases so different models can be tested against the real OSSOS detections.
This simulator uses a synthetic model to generate objects in our model twotino population and then determines whether the characterized survey could have detected each object based on detection efficiency and other parameters. 
See \citet{pet11} and \citet{law18} for an introduction and discussion of this kind of survey simulator.

Part of the reason this approach is so important is that both the location of the blocks relative to Neptune and their relative depths have a very strong effect on the orbital elements of the detections. 
This produces very non-uniform sensitivity to the current resonant angle and the body's long-term libration amplitude, an effect which is quantified below. For the present moment, we just point out that because $\lambda = \varpi + M$, where $M$ is the mean longitude,
\begin{equation}
\phit = 2 M + \varpi - \lambda_N .
\end{equation}
\noindent
Because there is a strong bias towards detection near perihelion where $M\approx 0$, most of the time the resonant particle's resonant amplitude at detection is roughly 
$\phit \approx \varpi - \lambda_N$, meaning that particles are usually detected with resonant arguments whose value is just the longitude difference ahead of Neptune of the observing location.
A characterized survey provides observing locations, depths, and a list of detections, but nearly as importantly the non-detection (or rare detection) of resonant objects in particular directions also provides valuable constraints on the intrinsic distribution of the resonance angles for the population.

\subsection{Twotino distribution in resonant phase space} \label{sec:phase}

\begin{figure*}[ht!]
\centering
\plotone{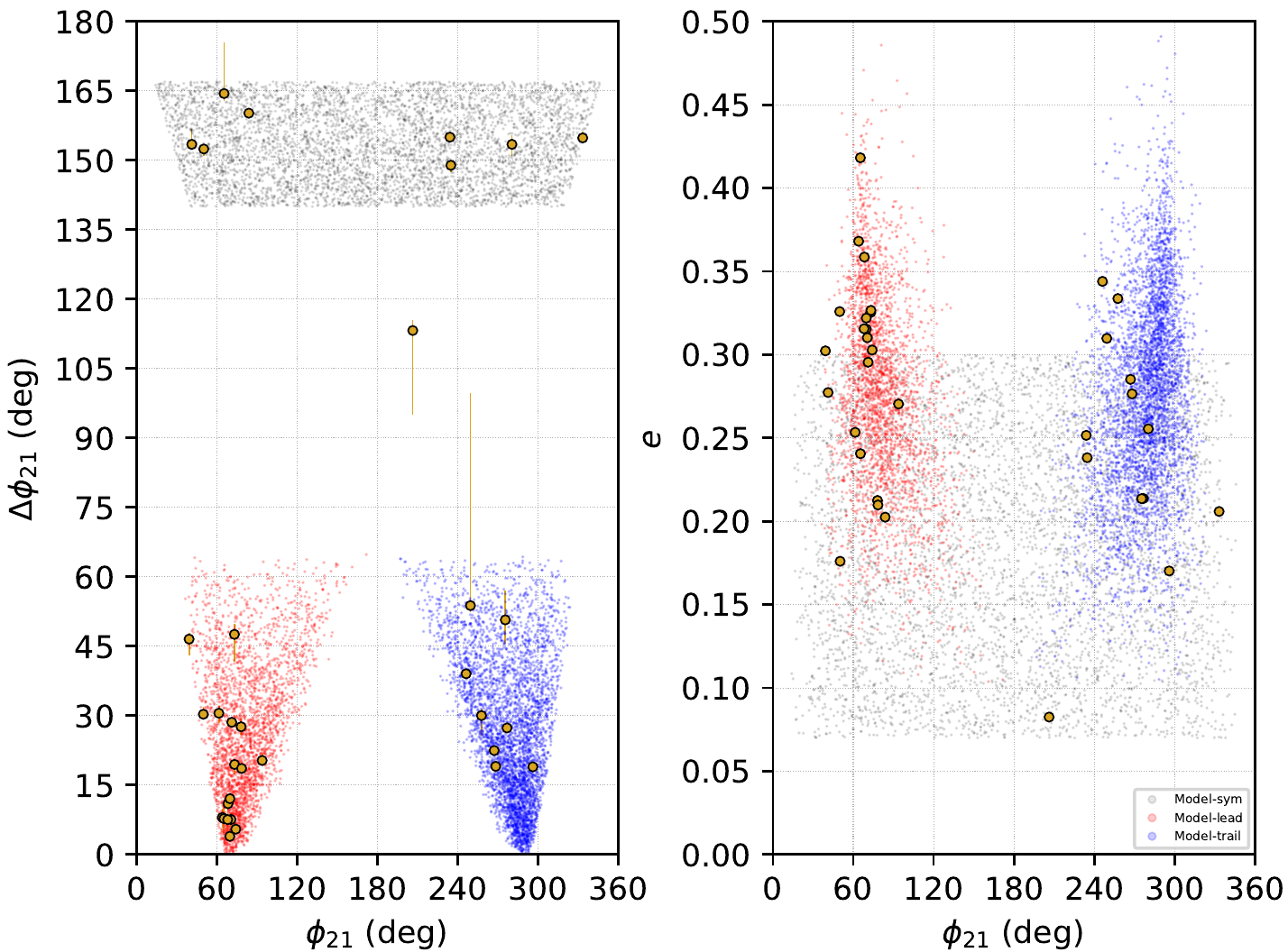}
\caption{The distribution of libration amplitude $\Delta\phit$ (left panel) and eccentricity (right panel) versus current resonant argument $\phit$.
The yellow dots indicate the OSSOS twotino detections (with libration amplitude uncertainties indicated as vertical lines in the left panel).
The gray, red, and blue dots show the model of the intrinsic twotino population for the symmetric, leading, and trailing islands respectively.
The object at $\phit\sim206\arcdeg$, $\Delta\phit\sim113\arcdeg$, and $e\simeq$ 0.08 is o3o33, which is currently transitioning from the symmetric island to the asymmetric island.
The object at $\phit\sim249\arcdeg$, $\Delta\phit\sim54\arcdeg$, and $e\simeq$ 0.31 is o5p121, which has larger uncertainties due to the existence of a few symmetric clones.}
\label{fig:fig_model_elibc}
\end{figure*}

\begin{figure*}[ht!]
\centering
\plotone{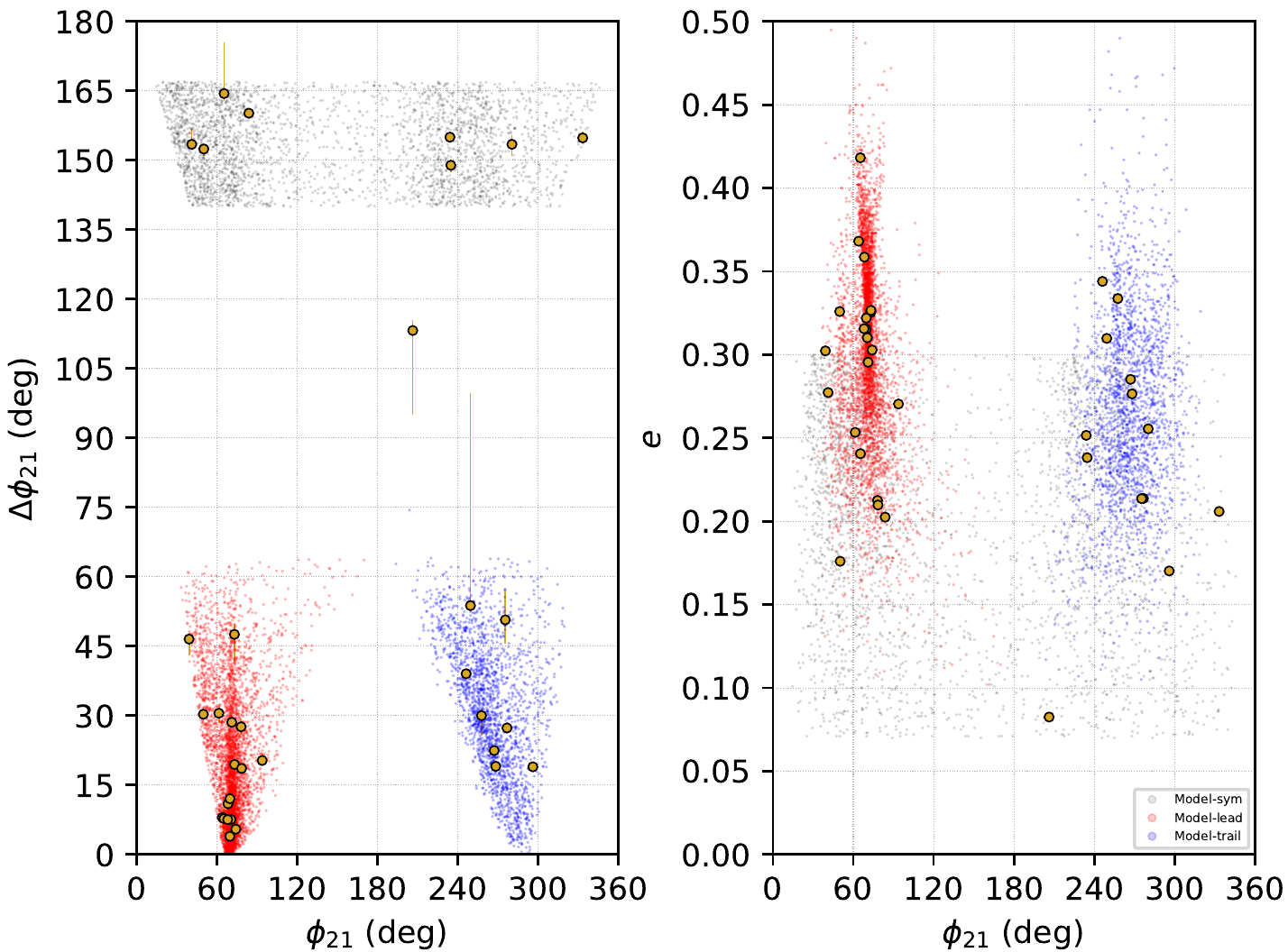}
\caption{Same as Figure~\ref{fig:fig_model_elibc}, but small dots now represent the twotinos from the model that are recovered (i.e., ``observed") after running them though the OSSOS survey simulator. The gray (symmetric) simulated detections show the survey's non-uniform sensitivity to twotinos with different values of $\phit$. The dense cluster of leading (red) dots near $\phit = 70\arcdeg$ are mostly model twotinos detected in the C/D observing block, this block had the greatest sensitivity to fainter objects and was located at the nearly optimal longitude for detecting leading twotinos.}
\label{fig:fig_tracked_elibc}
\end{figure*}

The construction of our synthetic model is based in part on the real detections and in part on the parameter space found to be occupied in previous studies of the dynamics of the 2:1 resonance. 
In modeling the CFEPS \citep{gla12} and the first quarter of the OSSOS twotino sample \citep{vol16}, an empirical model was adopted, which uniformly filled the stable phase space in eccentricity and libration amplitude identified in models of the current 2:1 resonance \citep{nes01,tis09}; this model was sufficient to describe the initially small sample of OSSOS twotinos.
With the full sample of \nt twotinos, this simplified, uniform model of the three libration islands can be statistically rejected.
To construct a more detailed model of the twotino population, we look to other models in the literature.
\citet{chi02} explored the distribution of libration centers $\langle\phit\rangle$ and amplitudes $\dphit$ at the end of planetary migration simulations; these distributions are shown in Figure~7 of \citet{chi02} and guide the modeling described in the following sections. 
In a similar, independent theoretical study that was more focused on the resonant trapping efficiency during migration as a function of initial inclination, \citet{li14} recovered similar distributions, although they showed that the regions of the ($\langle\phit\rangle$, $\dphit$) phase space inhabited today should depend on the inclinations of the initial TNOs before resonant trapping. 
Restricting the simulations of \citet{li14} to near-planar TNO initial conditions results in a very similar expected phase space distribution to that of \citet{chi02}.
In addition, \citet{mal18} provided the boundaries of twotino stability, which could be a guide to exploring ($a$, $e$) phase space.
Here we detail the distributions of intrinsic orbital parameters in our model, as guided by these theoretical models.
Our chosen range of parameters, and the distribution of the twotinos inside those ranges, are statistically not rejectable by the real observed twotino population, when biased by the survey simulator (see Figure~\ref{fig:fig_model_elibc} and \ref{fig:fig_tracked_elibc}).

\begin{figure}[ht!]
\centering
\plotone{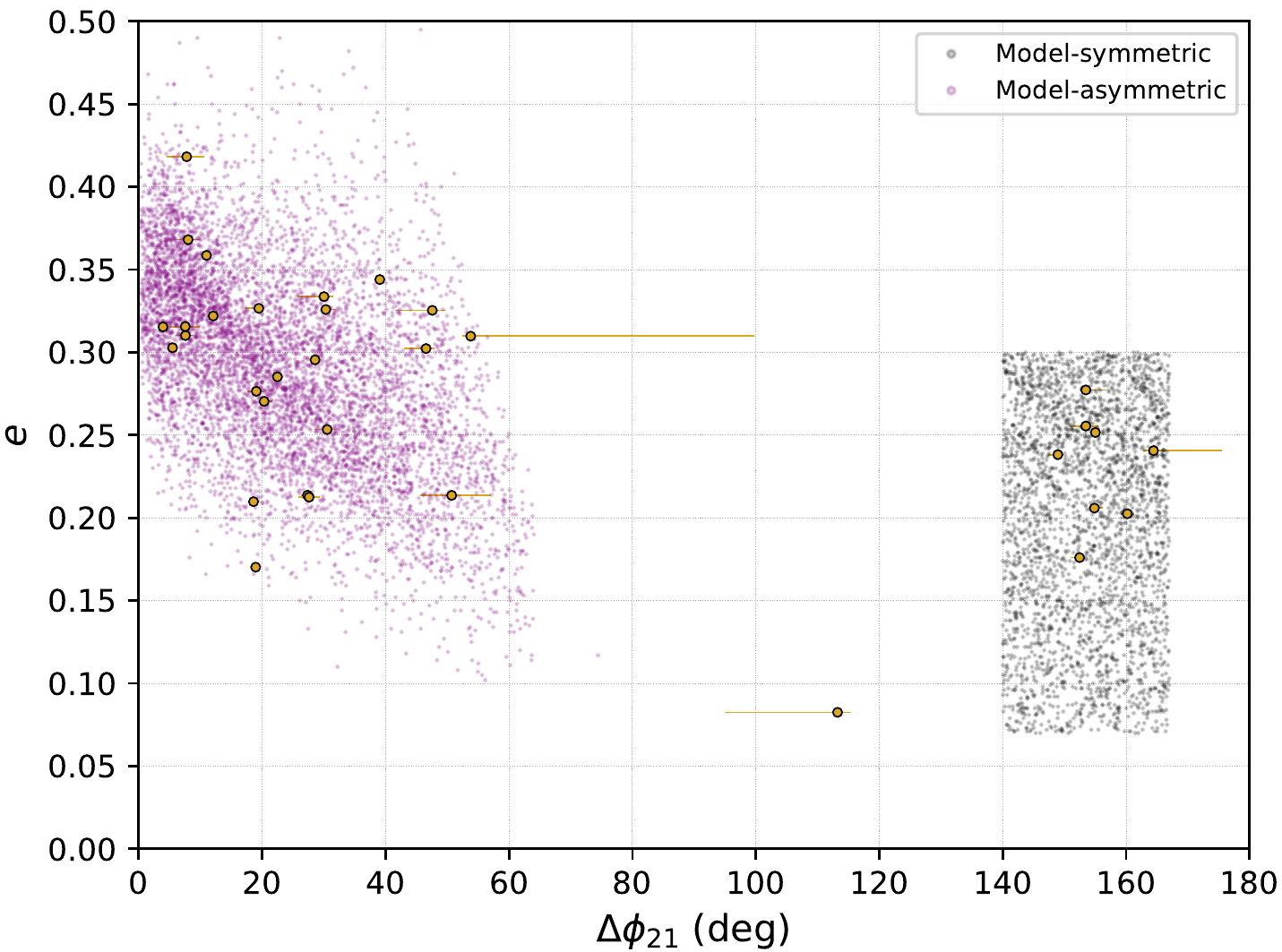}
\caption{The distribution of eccentricity and libration amplitude for the real OSSOS twotino detections (yellow dots) and the simulated detections from the twotino model (small purple and gray dots) after running it through the OSSOS Survey Simulator. As elsewhere, the island-changing o3o33 with $\Delta\phit\sim113\arcdeg$ is not part of our modeling.}
\label{fig:fig_tracked_eresamp}
\end{figure}

\subsubsection{Symmetric Twotino model} \label{sec:sym_model}

The symmetric twotino population is relatively simple to model.
All such objects have a libration center $\langle\phit\rangle=180^\circ$, and we assign a libration amplitude $\dphit$ randomly from a uniform distribution over the range 140 to 167 degrees; the amplitude range is based on the range showing 4~Gyr orbital stability in \citet{li14}.
Because we impose a sawtooth wave time evolution for $\phit$ (see Section~\ref{sec:dyna}, based on this being a better representation of the forward time evolution of detected OSSOS symmetric TNOs than a sinusoidal time history), this generates the uniform distribution in $\phit$ shown in the left panel of 
Figure~\ref{fig:fig_model_elibc}.
The orbital eccentricity $e$ for symmetric librators is modeled as uniform from $e = 0.07$ to 0.3.
These choices, when biased by the survey simulator, yield an acceptable match to the symmetric twotinos detected by OSSOS (Figure~\ref{fig:fig_tracked_elibc}).

We note that one of the nine symmetric librators detected by OSSOS is not described well by this model, because it has a rather low libration amplitude (o3o33/2013 JJ$_{64}$ with $\dphit \approx 113^\circ$);
in the known phase space structure of the resonance, such a small libration amplitude can only exist if the orbital eccentricity is small, and indeed with $e=0.08$, this is the lowest eccentricity object in the sample.
It is also the case that the libration behavior of o3o33 is not strictly symmetric on timescales longer than a few libration cycles. In the 10 Myr integrations performed for the dynamical classification, clones of o3o33 switch between the small amplitude symmetric libration described above and even smaller amplitude asymmetric libration around both the leading and trailing islands. Its behavior on timescales longer than $\sim1$~Myr is very chaotic, which is expected for very low eccentricity objects in the twotino population \citep[see, e.g.,][]{nes01}.
Given that only one such object was found by OSSOS, we have elected not to do the complicated modelling of this particularly chaotic portion of the symmetric resonant phase space.
Thus, all our symmetric population estimates concern the $0.07 < e < 0.3$ and $140\arcdeg < \dphit < 167\arcdeg$ portion of the phase space that essentially all other known symmetric twotinos inhabit, including the eight remaining OSSOS detections (Figure~\ref{fig:fig_model_elibc}, left panel).

\subsubsection{Asymmetric Twotino Population Model} \label{sec:asym_model}

Modeling the twotinos in the asymmetric islands is much more involved.
The structure of the resonance's asymmetric islands is not simple to represent in an analytic fashion, so we instead base our model on empirical relationships in the literature derived from numerical studies of these islands.
Note that we follow the same procedure for both asymmetric islands, except the determination of island at last step.
Our algorithm for generating an asymmetric twotino in our model is as follows:

\begin{figure*}[ht!]
\centering
\plotone{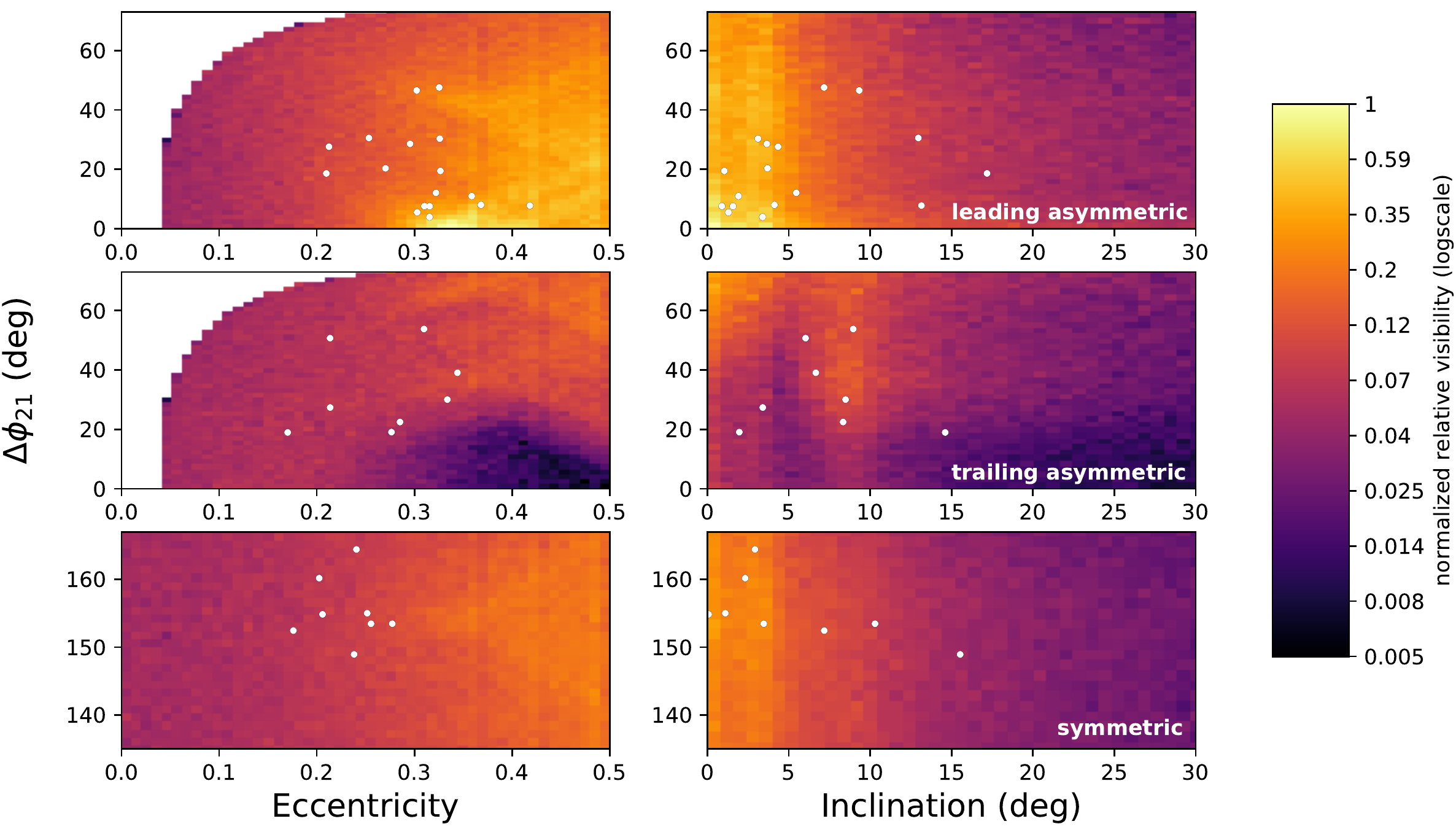}
\caption{Estimated visibility map for the phase space of the 2:1 resonance in OSSOS. To generate this map, we assume a \emph{uniform underlying distribution} in $e$, $i$, and $\dphit$ within the allowed resonant phase space for each libration island and a slope of $\alpha$ = 0.6 for the $H$ distribution.
The color scale is log(normalized visibility) where normalized visibility = 1 occurs for the ($e$, $\dphit$) bin with the most simulated detections across all three libration islands when the model is passed through the OSSOS survey simulator in the left-hand plots and for the ($i$, $\dphit$) bin with the most modeled detections in the right-hand plots.
Note that the white regions in the middle-left and top-left panels are just the phase spaces which exceed the maximum $\dphit$ in \citet{nes01}, and thus have no generated twotinos.
}
\label{fig:fig_visibility}
\end{figure*}

\begin{enumerate}

\item Choose an eccentricity.

The $e$ distribution turns out to be satisfactorily modeled as a Gaussian with mean ($e_\mathrm{c}$) at $e=0.275$ and a width ($e_\mathrm{w}$) of 0.06. We truncate the outlying tails of this Gaussian distribution to force values of $e$ to be in the range 0--0.53; the upper limit represents the eccentricity where an object's perihelion distance allows strong interactions with Uranus and would thus be unstable.
This intrinsic $e$ distribution can be seen in the right panel of Figure~\ref{fig:fig_model_elibc}, with the expected tendency to detect the higher-$e$ members evident in Figure~\ref{fig:fig_tracked_elibc}'s right panel.

\item Choose the libration center $\langle\phit\rangle$.

To get an acceptable match, we converged on the following empirical algorithm.
The libration center $\langle\phit\rangle$ is highly correlated with $e$ \citep[see Figure~10c of][]{chi02}. To generate values of $\langle\phit\rangle$ from their model distribution, 
we use following polynomial functions to fit Figure~10c of \citet{chi02}. We first calculate a minimum value from the polynomial function
\begin{equation}
\langle\phit\rangle_\mathrm{low} = 133.2 - 315.3 e + 377.1 e^2 - 0.4916/e.
\end{equation}
To account for the fact that there is some variation in $\langle\phit\rangle$ even for a single value of $e$,
we then pick a random number $f$ on (0,1) and weight the libration center towards the low-amplitude edge via
\begin{equation}
\langle\phit\rangle = \langle\phit\rangle_\mathrm{low} + 20\times \left[ 1 - \sin\left(\frac{f\pi}{2}\right)\right].
\end{equation}
This tendency for asymmetric twotinos to be concentrated towards the lowest libration center allowed for a given $e$ was required to get a satisfactory match and is thus an aspect of the real distribution.
\item Choose the libration amplitude $\dphit$.

We bound the lower and upper limits of the libration amplitude distribution as
\begin{eqnarray}
{\dphit}_\mathrm{min} & = & 79.03 
     \times \exp \left[ -
     \frac{(\langle\phit\rangle - 121.3)^2 }{ 2 \times 15.51^2} \right]\\
{\dphit}_\mathrm{max} & = & 9.10\langle\phit\rangle - 0.04425\langle\phit\rangle^2 -  0.0884/\langle\phit\rangle - 404
\end{eqnarray}
based on the boundaries of the asymmetric islands in Figure~7 of \citet{chi02}.
We then generate the amplitude $\dphit$ between
these limits, weighting lower amplitudes more heavily 
using a linear distribution with a slope of -0.75. 

\item Choose the current value of the resonant argument.

We choose $\phit$ from a uniform distribution between its limits by choosing a random phase $R$ in the interval \mbox{(-1, 1)}, and then setting
\begin{equation}
\phit = \langle\phit\rangle + R \times \dphit.
\end{equation}

\item Choose leading or trailing island.

The model defines a fraction, $0 < f_\mathrm{L} < 1$, of the asymmetric twotinos that are in the leading island. To assign a libration island, a uniform random number in this range is drawn and if it is larger than $f_\mathrm{L}$ the object is converted to a trailing twotino by the simple transformation
\begin{equation}
\langle\phit\rangle \rightarrow 360^\circ - \langle\phit\rangle.
\end{equation}
We treat $f_\mathrm{L}$ as a free parameter, and we vary it in order to match the observed, debiased distribution (see Section~\ref{sec:fraction}).
\end{enumerate}

Biasing the model twotino distribution described above using the OSSOS survey simulator yields the distribution shown in
Figure~\ref{fig:fig_tracked_eresamp}, which is compared statistically to the real detections in Section~\ref{sec:verify}. 
OSSOS detected 17~leading and 8~trailing asymmetric twotinos.
The dominant reason for the roughly factor of two enhancement in leading island detections is not that they are intrinsically more numerous.
As a comparison of Figure~\ref{fig:fig_model_elibc} and \ref{fig:fig_tracked_elibc} already shows, the OSSOS blocks that pointed in the direction of the leading asymmetric librators achieved fainter limiting magnitudes than those pointed toward the trailing librators and thus yield the majority of the expected detections, even for a population with equal numbers of leading and trailing twotinos.
This is most clearly illustrated in Figure~\ref{fig:fig_visibility}, which shows how sensitive OSSOS was to the phase space within each libration island. 
This `visibility map' is constructed by uniformly filling the phase space of the resonance in $e$, $i$, and $\dphit$ within the limits shown in the figure and plotting the relative distribution of the resulting simulated detections.

The resulting Figure~\ref{fig:fig_visibility} nicely illustrates the strong biases 
induced by the survey's pointing and the flux bias.
For example, there is a very strong bias favoring the detection of $e>0.3$ symmetric twotinos which has only weak dependence on libration amplitude (because the latter are all large and thus high-amplitude twotinos could be present in any OSSOS block);  the lack of such detections
by OSSOS proves they are unstable or scarcely populated, and motivated us to place an upper limit. The Figure~\ref{fig:fig_visibility} also
shows the clear bias toward the leading asymmetric island (top panels) compared to the trailing island (middle panels) and required the upper eccentricity cutoff of 0.3 in the model.
The difference in the biases for asymmetric twotinos is striking, which is caused by
the location of Neptune relative to the galactic plane during the OSSOS epochs.
For leading asymmetric twotinos, the blocks from RA=0h to 3h mean that eccentric objects at
perihelion will be close and bright, and the C/D block (Figure~\ref{fig:ossos_top_view})
is perfectly placed for the detection of very low libration amplitude objects; there
is thus a strong concentration of objects where bias is high, but also a need to have
the intrinsic $e$ distribution to decline sharply above 0.4.
In contrast, for the trailing blocks, low libration amplitude twotinos would have
perhelia longitudes in the galactic plane and if they have large $e$ they will be
at large $d$ when in the blocks between RA=13h and 16h; for larger libration
amplitudes some better detectability develops because the periehlia for some
of the objects can librate over to first M and O blocks and then E and P.
The trailing twotinos again suggest $e>0.4$ is rare or absent.
Lastly, the right column shows the expected general bias towards the detection
of lower-$i$ twotinos; the fact that there are detections with $i>10^\circ$ implies
that the true inclination distribution has a certain number of large-$i$
members.
In addition, the leading (and symmetric) panels shows that placing M block
to cover latitudes from 6 to 10 degrees has an important effect at allowing
the detection of $i>5^\circ$ to better constrain the inclination distribution.
Figure~\ref{fig:fig_visibility} was intended to illustrate the general biases present and how complicated they can be; the underlying distributions are NOT uniform in these
quantities, and we will below validate our parameterizations of the non-uniform
orbital element distribution.
Only with the ability to precisely measure the detection biases can one reliably
try to extract the true orbital element distributions.  
OSSOS was {\it designed} precisely to do this.

\begin{figure*}[ht!]
\centering
\plottwo{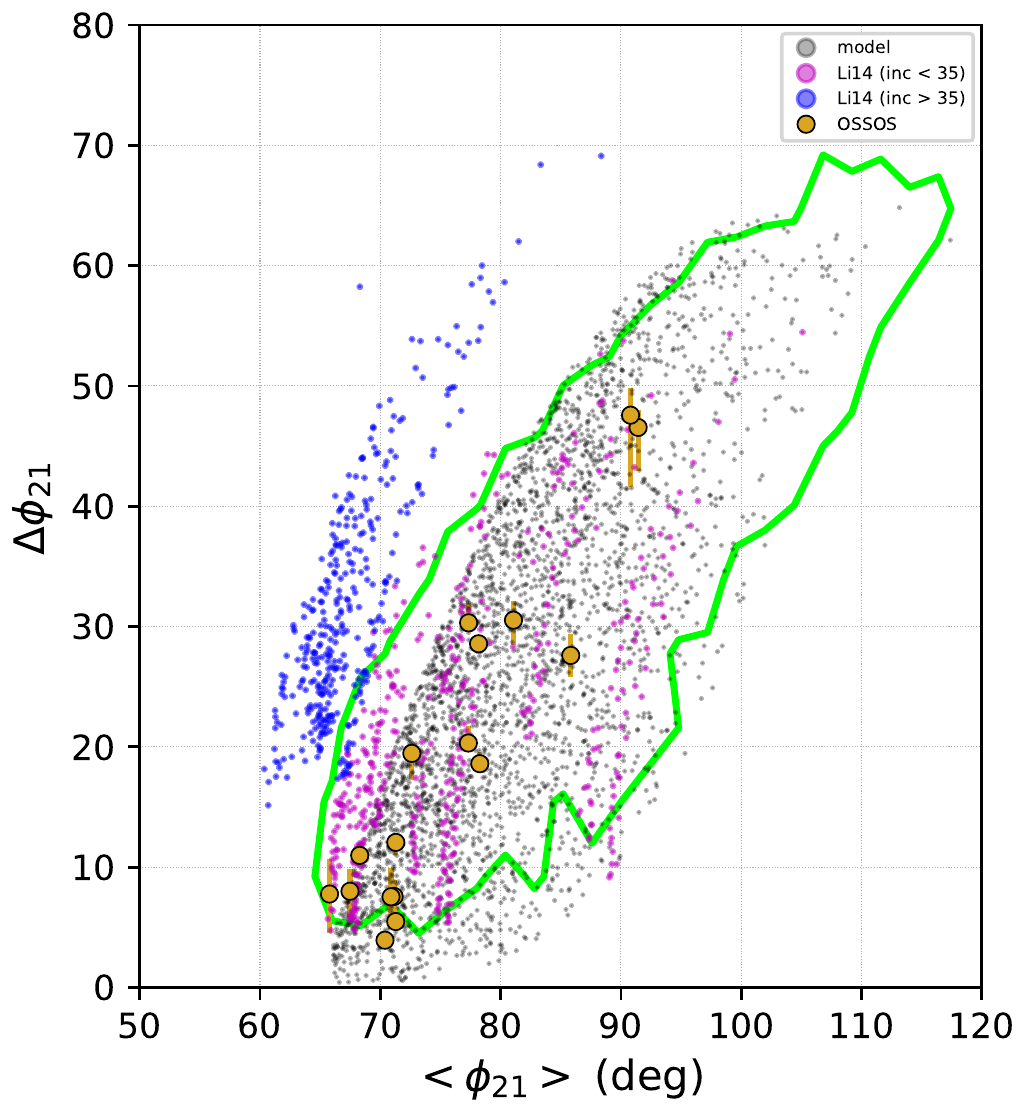}{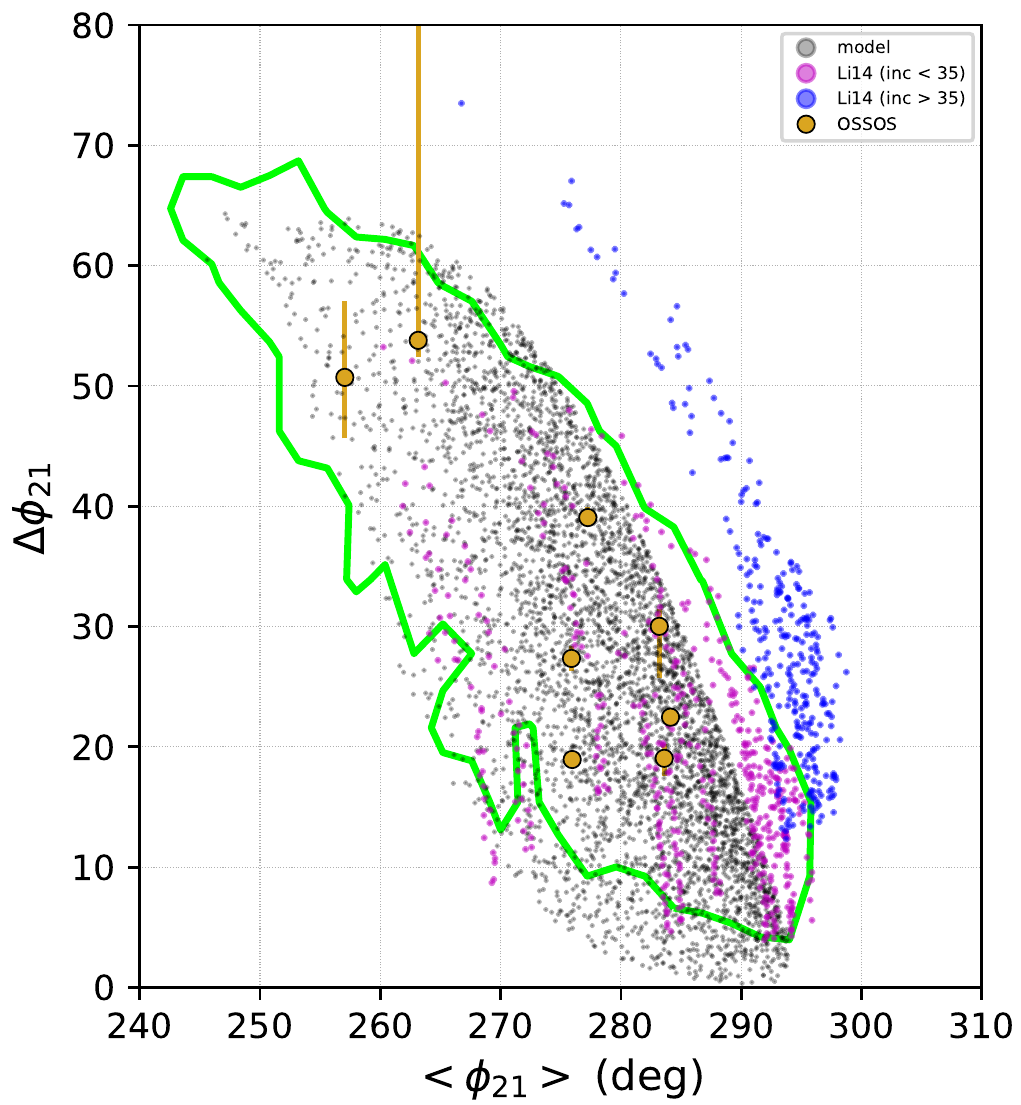}
\caption{The distribution of libration centers and libration amplitudes for the leading (left panel) and trailing (right panel) islands.
Yellow points are the OSSOS twotino detections (which are a biased set).
The black points show our favored model distribution (unbiased), based on the work of \citet{chi02} and adjusted to match the OSSOS sample in Section~\ref{sec:verify}.
The magenta and blue points show the results of a capture study \citep{li14} for initial orbital inclinations of $i<35^\circ$ and $i>35^\circ$ respectively; the low-$i$ results are a better match to the OSSOS twotino population (all of which have $i<35^\circ$) and are consistent with the distribution simulated in \citet{chi02}, as expected.
The contours indicate the stable phase space of the twotino population based on the surviving particles after a 10~Myr numerical simulation (see text for dicussion). 
}
\label{fig:fig_model_libcresamp}
\end{figure*}

For a final perspective of the complexity, we implemented a numerical simulation using MERCURY \citep{cha99} to obtain a stability map of the boundaries in the libration center and width space.
We randomly generated 40000 synthetic particles with $47.2$~au $\leqslant a \leqslant 48.5$~au, $0\arcdeg \leqslant i \leqslant 20\arcdeg$ and $0 \leqslant e \leqslant 0.41$. 
In order to generate orientation angles that gave in intially librating asymmetric twotino orbit,
the resonant arguments and amplitudes are randomly chosen within the ranges described in section \ref{sec:dyna}. Each libration island had an equal number of particles. Using Equation \ref{eq:resang}, the argument of perihelion is calculated with random ascending node, random true anomaly and resonant arguments/amplitudes chosen above. We identified 8134 surviving particles librating as twotinos~at the end of a 1~Gyr simulation. (The surviving particles were identified as twotinos if the 2:1 resonant argument librated over the last 5 Myr). We note that a few of these particles are in the phase of island switching like o3o33 described in Section \ref{sec:sym_model}.
The libration centers and amplitudes of the surviving particles in this simulation are bounded by the contours in Figure~\ref{fig:fig_model_libcresamp}. 
This stability map of the synthetic particles is relatively consistent with the measured distribution of the OSSOS twotinos, our unbiased model, and the simulations of \citet{chi02} and \citet{li14} (assuming initial conditions of low inclination). 
The biggest exception is the abundance of low-amplitude leading librators.
This discrepancy likely results from the uniform initial distribution of orbital elements in the 
numerical study; initial conditions that give $\dphit<10^\circ$ are confined to
a small portion of phase space and are under-represented whereas the detection biases greatly
favor their discovery in OSSOS, as just discussed.
A second, smaller, discrepancy is that the numerical study retains 
more moderate to large libration amplitude twotinos. 
Most of integrated particles in this region frequently transit between the symmetric island and an asymmetric island, which may artificially inflate the measurement of their asymmetric libration amplitude. This discrepancy may also be due to our initial particle distribution of particles, but in the opposite way (over-representing large amplitude librators compared to how the Solar System populated the resonances).

\begin{figure}[ht!]
\centering
\plotone{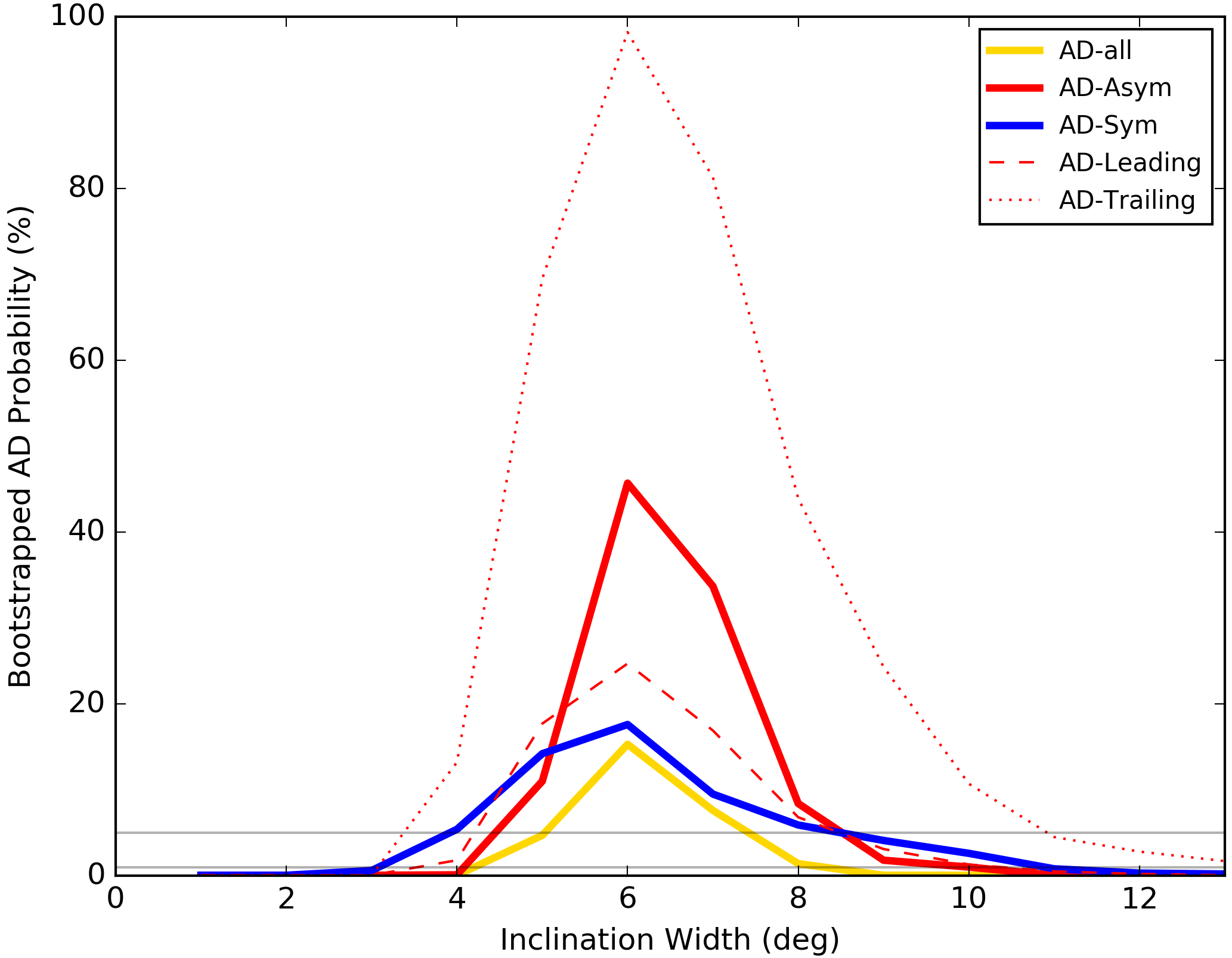}
\caption{The bootstrapped Anderson-Darling test distribution vs width of the inclination distribution $\sigma_\mathrm{i}$ for the leading asymmetric (dashed line), trailing asymmetric (dotted line), all asymmetric (red line), symmetric (blue line), and all twotino (yellow line). A value of $\sigma_\mathrm{i}=6\arcdeg$ has the lowest rejectability for each set of twotinos, independently, and for the entire ensemble. The horizontal lines at 5\% and 1\% show models rejected at moderate and high confidence.
} 
\label{fig:fig_inc_prob}
\end{figure}

\section{Validation of the Model} \label{sec:verify}

The validity of our empirical twotino model is established by comparing the simulated detections with the real OSSOS twotino detections. 
In our comparison, we consider the given orbital elements separately as a one-dimensional cumulative distributions. 
We use the bootstrapped Anderson–Darling (AD) test to compare the simulated 1-D distributions to the observed distributions. The AD statistic for the N real twotinos is computed by randomly drawing subsamples of N synthetic detections from the biased detections and bootstrapping the AD statistic for these subsamples; we choose to reject models if the AD statistic for N real twotinos is below the 5\% confidence level.
The verification includes $e$ ($e_\mathrm{c}$ and $e_\mathrm{w}$), $i$, $d$, $H$, $\phit$, $\langle\phit\rangle$, and $\dphit$, to identify the rejectability of model parameters.

The validation of the models was done in following steps:

\begin{enumerate}
\item \textbf{Libration centers and amplitudes for asymmetric twotinos:} The range of the current resonance angles, $\phit$, and libration amplitudes, $\dphit$, generated within the model was first confirmed to encompass the values of the known detections.
After that, the boundaries for these distributions were fixed.
Figures~\ref{fig:fig_model_elibc} through~\ref{fig:fig_tracked_eresamp} show that the model covers the known detections, excepting the one unstable low-$e$ symmetric object mentioned
in Section~\ref{sec:sym_model} (o3o33) that we discard for the purposes of this study.

\item \textbf{Inclination:} The inclination distribution was then adjusted.
TNO inclination distributions are acceptably represented by a functional form of sin($i$) times a Gaussian distribution of width $\sigma_\mathrm{i}$ \citep{bro01, gul10, gla12}.
We varied this width freely to determine what range is not 
rejectable using the bootstrapped AD test.
Figure~\ref{fig:fig_inc_prob} shows the results of this statistical analysis of the twotinos inclination distributions in the three libration islands. 
In all cases, the OSSOS detections can be acceptably modeled by an intrinsic inclination distribution with a
 width of $\sigma_\mathrm{i}=6\arcdeg$, with the 95\% confidence interval being roughly \mbox{4$\arcdeg$ -- 8$\arcdeg$} for different libration islands. 
Note that, although there are fewer detections, the symmetric population independently supports this same inclination width range, as does each asymmetric island when tested independently.
As a possible source of twotinos, the scattering TNOs could transiently stuck in 2:1 resonance, and occupy a significant ratio of symmetric twotino \citep{yu18}. The peak of the inclination distribution of the transient-sticking twotinos only slightly shifts up by $\sim1^\circ$, i.e. the transient-sticking twotinos have a similar hot inclination distribution as the scattering TNOs. However, the OSSOS symmetric twotinos prefer a colder inclination distribution, which indicate that the twotinos probably originate from the primordial disk rather than the scattering disk.
For comparison, the inclination distribution width of the plutinos was estimated by \citet{vol16} to have a width of $8\arcdeg \leqslant \sigma_\mathrm{i} \leqslant 21\arcdeg$ at 95\% confidence level, and by \citet{ale16} to have a width of $11\arcdeg \leqslant \sigma_\mathrm{i} \leqslant 21\arcdeg$ at 95\% confidence level; this range of widths is essentally non-overlapping for the twotino inclination distribution in this study.
The intrinsic inclination distribution of plutinos and twotinos are thus different, which must be a result of how their resonances were populated and eroded.
We thus fix $\sigma_\mathrm{i}=6\arcdeg$ for our nominal twotino model that will be used to estimate the total population and ratio of asymmetric islands.
The quality of the match between simulated and real OSSOS detections for this modeled inclination distribution is shown later in Figure~\ref{fig:fig_cum_dieh}. The cosmogonic implications of this narrow width are discussed in Section~\ref{sec:dis}.

\begin{figure*}[ht!]
\centering
\plotone{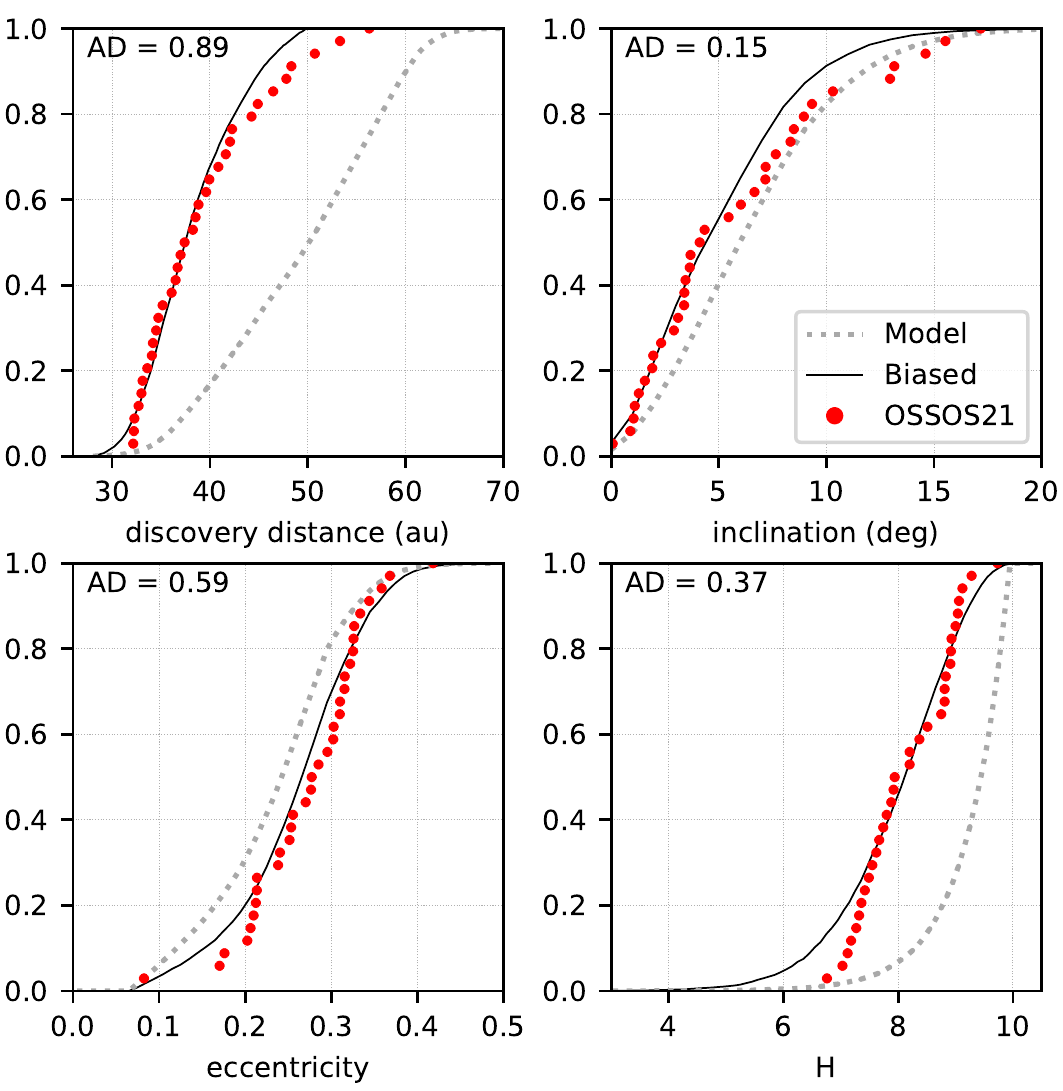}
\caption{Cumulative plots of orbital parameters for discovery distance ($d$), orbital inclination ($i$), orbital eccentricity ($e$), and the absolute magnitude $H$ distribution for the full model of all OSSOS twotinos (symmetric and asymmetric twotinos).
The gray dotted line shows the intrinsic (model) distribution for the parameter; the black line indicates the biased results produced by the OSSOS survey simulator. 
The red dots give the distribution of the real twotinos detected by OSSOS.
The real detections are statistically consistent with being drawn from the biased distribution.
}
\label{fig:fig_cum_dieh}
\end{figure*}

\item \textbf{Parameters for symmetric twotinos:} The validation of the symmetric twotinos is relatively straightforward. As mentioned in \ref{sec:sym_model}, we adopt a uniform distribution in an appropriate range for $\phit$, $\dphit$, and $e$. 
This simple model is a statistically acceptable match to the OSSOS detections. The validation of $e$, $i$, $d$, $H$, $\phit$, $\langle\phit\rangle$, and $\dphit$ distribution, is non-rejectable on the condition of 95\% confidence level. Based on only eight symmetric twotinos in OSSOS sample, we avoid overinterpretation of details and are mostly interested in the population size. 

\item \textbf{Size distribution and eccentricity for asymmetric twotinos:} We adopt a single exponential $H$-magnitude distribution because it turns out to not be rejectable, although
the feature near $H$ = 8.1–-8.5 in the $H$ distribution of Figure~\ref{fig:fig_cum_dieh} could be a hint of a knee or divot H-distribution \citep{law18b}.
We assume the magnitude distribution is the same for twotinos in all three islands.
We then vary the logarithmic slope, $\alpha$, of the cumulative absolute magnitude exponential distribution $N(<H_r) \propto 10^{\alpha H_r}$ and, for the asymmetric twotinos we vary the Gaussian center $e_c$ and width $e_w$ of the eccentricity distribution.
These three parameters are strongly coupled because of the observational biases, and we thus consider them together in a single statistical analysis.
(In contrast, the symmetric twotinos were acceptably modeled with a simple, uniform eccentricity distribution.) 
We then repeat the analysis of \citet{vol16} (their Figure 4), computing the bootstrapped probability
that the real sample's Anderson-Darling (AD) statistic would be more discrepant than equally-sized 
sub-sample from a model defined by the ($\alpha$, $e_c$, $e_w$) triplet.
We assign a model a rejectability that is the worst of the three AD probabilities for the
$e$, $d$, and $H$ cumulative distributions
; we found that this
choice did a better job of ruling out a model which is clearly discrepant in one
variable than did the 'summed AD statistic' used previously.
Rejectability maps of these three parameters are shown in Figure~\ref{fig:fig_adsum_grid}.
The least rejectable trio of values is $\alpha = 0.6$, $e_\mathrm{c} = 0.275$, and $e_\mathrm{w} = 0.06$, and we adopt these parameters as our nominal model; the analysis shows that the data suggest an
asymmetric eccentricity distribution that is strongly peaked near $e_c\simeq0.28$ with a 
narrow width of only about 0.06, and that such a model provides and entirely acceptable
match to the data (with a worst-of-three AD probability as high as 0.58).
To 95\% confidence,
we cannot rule out the case of lower $e_\mathrm{c}$ and large $e_\mathrm{w}$ (that is, a wide Gaussian centered at low $e$) due to the detection bias against low-$e$ twotinos, but such models produce
worse matches to the eccentricity distribution.
An exponential with a single value of $\alpha$ (i.e. a size distribution without a break or divot)
satisfactorily represents the data, but is the worst of the matches (Figure~\ref{fig:fig_cum_dieh}).
We also individually tested the size distribution with a few published size distributions: (1) knees in \citet{fra14}, (2) the knee in \citet{law18b}, and (3) the divot in \citet{law18b}. The knee and divot in \citet{law18b} (based on the scattering population) are not rejectable and provide highter AD values then that in this study, whereas both knees for the hot population in \citet{fra14} are nearly rejectable (at 93\% confidence level).
The H-magnitude distribution is likely the first thing that will be modified if a future well-characterized survey can more than double the OSSOS sample, 
and will likely then formally reject the 
single exponential.

\begin{figure*}[ht!]
\centering
\plotone{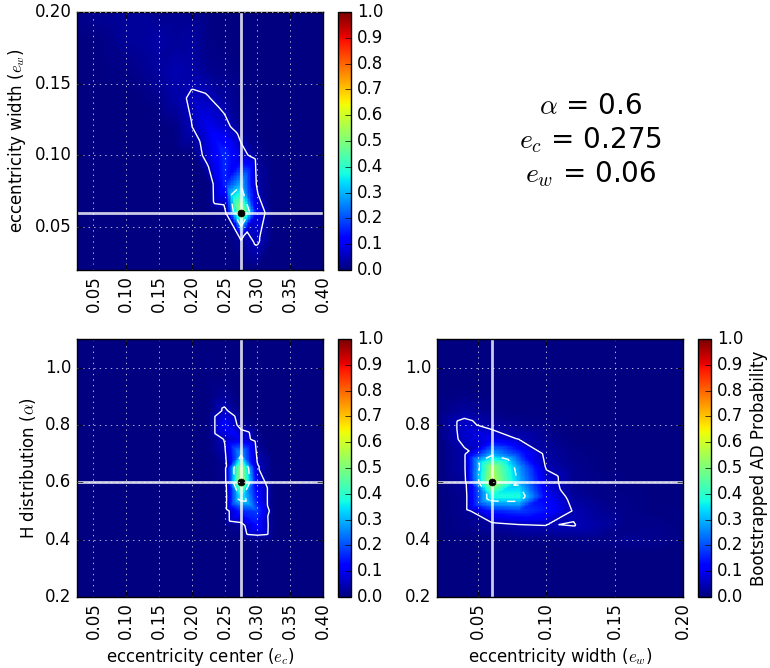}
\caption{Bootstrapped AD probability maps for a range of values of $\alpha$, $e_\mathrm{c}$, and $e_\mathrm{w}$ in the intrinsic asymmetric twotino model.
The area outside of the white contours indicates a rejected value of the worst AD statistic at $>$95\% confidence level (solid) or 68\% confidence level (dashed). The most non-rejectable model parameters are shown by the black dots and are stated at the top right. Each panel is a two-dimensional cut of our three-dimensional parameter space search. For each panel, we fix one parameter to its best value and show the AD rejectability map for the other two parameters.}
\label{fig:fig_adsum_grid}
\end{figure*}

\item \textbf{Fraction of all twotinos in the symmetric island, and the fraction of the asymmetric twotinos in the leading island:} Finally, we estimate the parameters $f_\mathrm{S}$ (the fraction of all twotinos which are symmetric librators) and
$f_\mathrm{L}$ (the fraction \emph{of the asymmetric twotinos} which are in the leading asymmetric island). Thus the fraction of all twotinos in the leading island is (1-$f_\mathrm{S}$)*$f_\mathrm{L}$. The OSSOS sample constrain that $0.2<f_\mathrm{S}<0.55$ and $0.2<f_\mathrm{L}<0.65$ at the 95\% confidence level. This analysis is described in detail in Section~\ref{sec:fraction}.

\end{enumerate}

\begin{figure*}[ht!]
\centering
\plotone{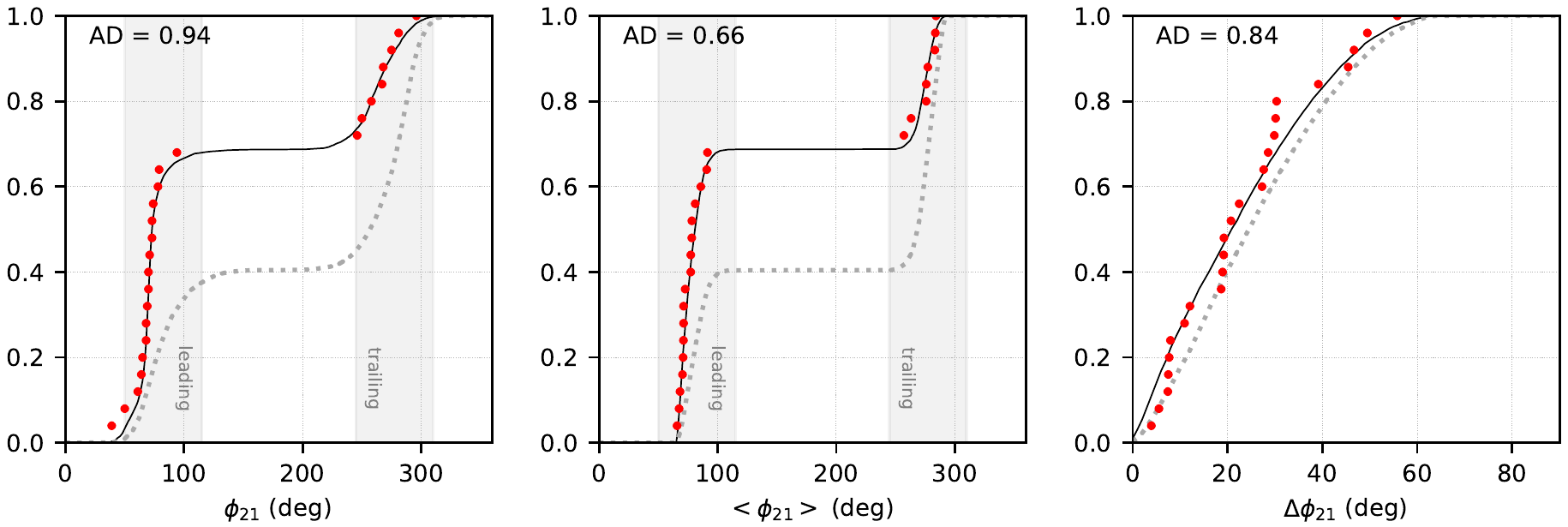}
\caption{Cumulative plots of the libration parameters for the two asymmetric islands (combined).
The red dots are the real detections, the gray dotted curves show the intrinsic empirical model, and the solid black curves indicate the simulated detections from the empirical model.
The left and middle panels illustrate the heavy detection biases present in the OSSOS twotino sample. In particular, the middle panel shows that the intrinsic 60/40 split between the trailing and leading island populations is heavily biased in the detected sample towards more detected leading twotino. This is due to the greater depth and favorable location of the fall blocks (see Figure~\ref{fig:ossos_top_view}); nearly 70\% of the detected asymmetric sample is thus found in the leading island.} 
\label{fig:fig_cum_phis}
\end{figure*}

Figures~\ref{fig:fig_cum_dieh} and \ref{fig:fig_cum_phis} show cumulative plots of various orbital parameters for our nominal model before and after being subjected to the OSSOS survey simulator compared to the real OSSOS twotinos. The real and modeled detections are non-rejectable matches. These certainly suffice to provide reasonable population estimates of the various twotino islands, which is our main goal.

\subsection{Population Estimates} \label{sec:pop}
To estimate the total twotino population in each libration island, we generate simulated twotinos from the models described above,
using the exponential absolute magnitude distribution with slope $\alpha = 0.6$, to a magnitude cutoff of $H_r < 10.0$.
This cutoff value covers the full range of OSSOS twotino $H_r$ magnitudes.
For each libration island, model twotinos are generated and run through the survey simulator until the number of tracked objects matches the number of detected twotinos. This test is repeated $10^4$~times for each island. The median number of simulated twotinos that must be generated to match the real number of detections is our estimate for the relevant twotino population; the distribution of number of simulated twotinos is used to calculate the uncertainty in our estimates.

\begin{deluxetable*}{llllll}
\tabletypesize{\scriptsize}
\tablecaption{Twotino Population Estimates \label{tab:pop}}
\tablehead{
\colhead{Libration island} & 
\colhead{$e$ distribution} &
\colhead{Median Pop} &
\colhead{Median Pop.} &
\colhead{Gladman 2012} &
\colhead{Volk 2016}\\
\colhead{}  & \colhead{}  &
\colhead{$H_r < 10$} &
\colhead{$H_r < 8.66$\tablenotemark{a}} &
\colhead{$H_g < 9.16$} &
\colhead{$H_g < 9.16$}}
\startdata
symmetric & uniform $e$ = 0.07–0.3   &  6500$^{-2900}_{+4300}$ & 1300$^{-600}_{+900}$&&\\ 
leading & $e_\mathrm{c}$ = 0.275, $e_\mathrm{w}$ = 0.06 &  4800$^{-1700}_{+2200}$ & 1600$^{-600}_{+600}$&&\\ 
trailing & $e_\mathrm{c}$ = 0.275, $e_\mathrm{w}$ = 0.06 &  7100$^{-3400}_{+5200}$ & 1500$^{-700}_{+1000}$&&\\ 
total&&18400$^{-4800}_{+7100}$&4400$^{-1100}_{+1500}$&3700$^{-2400}_{+4400}$&4000$^{-2000}_{+2500}$\\
\enddata
\tablenotetext{a}{The limit of $H_r$ = 8.66 is equivalent to the limit of $H_g$ = 9.16 used in \citet{gla12} and \citet{vol16}, assuming $g - r = 0.5$.}
\end{deluxetable*}


Table \ref{tab:pop} lists the median number of twotinos consistent with the real OSSOS detections in each libration island based on our nominal model as well as the 95\% confidence ranges for these population estimates. 
Note that the population estimate for the trailing island is more uncertain than that for the leading island due to the survey being less sensitive to twotinos in the trailing island (Figure~\ref{fig:fig_visibility}).
For comparison with the earlier results of \citet{gla12} and \citet{vol16}, these population estimates down to $H_r=8.66$ are also listed in Table \ref{tab:pop}. 
\citet{gla12} and \citet{vol16} make use of CFEPS data which was acquired in $g$ band, and set a faintest magnitude of $H_g = 9.16$. To convert to $r$, we assume $g - r = 0.5$ (thus the last three columns are comparable), which were used in \citet{vol16} and approximately consistent with the mean twotino color in \citet{mar19}.
Although the small number of twotinos in the two earlier studies made it impossible to usefully constrain the populations of the individual libration islands, their reported estimates of the total twotino population are in statistical agreement with our results, with OSSOS roughly halving the estimated uncertainty.

\subsection{Relative Populations of the Different Islands} \label{sec:fraction}
\begin{figure*}[ht!]
\centering
\plotone{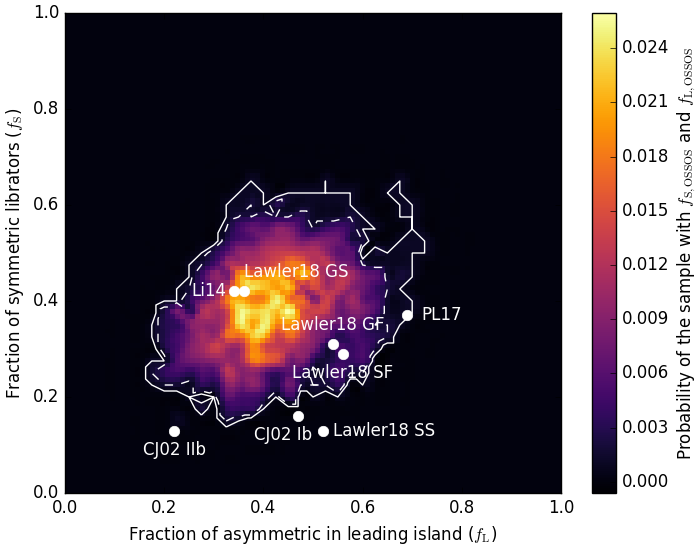}
\caption{The probability map of the fraction of symmetric island twotinos ($f_\mathrm{S}$), and fraction of the asymmetric twotinos in the leading island ($f_\mathrm{L}$), constrained by the observed population fractions of the symmetric and leading asymmetric islands ($f_\mathrm{S,OSSOS}$, $f_\mathrm{L,OSSOS}$) (i.e., the number of detections in each island should be same as the ratio in the OSSOS sample). The rejected regions for this probability distributions are indicated with solid white curves (99\% confidence level) and dashed white curves (95\% confidence). 
Note that the absolute probabilities (color map) are low only because the cells have very high resolution;
the contours correctly enclose the normalized probabilities.
The $f_\mathrm{S}$ and $f_\mathrm{L}$ from \citet{chi02}, \citet{li14}, \citet{pike17b}, and \citet{law18c} are overplotted with dots and labeled. 
}
\label{fig:fig_flts}
\end{figure*}

The population ratio of the asymmetric islands in the current epoch could provide clues to Neptune's migration rate. The fraction of symmetric twotinos will also constrain theoretical models, but has little discussion in the literature. 
For statistical analysis of these populations, we repeated the work shown in Figure~10 of \citet{vol16}, which only used nine twotinos (four from 13AO and 13AE block of OSSOS, and five from CFEPS).
For each candidate value of $f_\mathrm{S}$ and $f_\mathrm{L}$ on a grid from 0-1 using steps of 0.025, we generate twotinos (using the the favored empirical model described above) until the survey simulator produces 10000 tracked twotino detections.
Then we randomly draw a sample of \nt from these simulated twotino detections and check if that sample matches the OSSOS observed symmetric fraction of 9/34 and the OSSOS observed leading asymmetric fraction of 17/25. 
This process was repeated 1000 times to calculate a probability distribution across the tested range of $f_\mathrm{S}$ and $f_\mathrm{L}$ (see Figure~\ref{fig:fig_flts}).

OSSOS provides the first meaningful quantitative bounds on the relative island fractions.
The OSSOS sample demands with high certainty that $0.2<f_\mathrm{S}<0.6$ and $0.2<f_\mathrm{L}<0.7$ at the 99\% confidence level, and that $0.2<f_\mathrm{S}<0.55$ and $0.2<f_\mathrm{L}<0.65$ at the 95\% confidence level.
The figure shows a mild preference for both $f_\mathrm{S}$ and $f_\mathrm{L}$ at 40\%, but a 50\% value for $f_\mathrm{L}$ is perfectly acceptable.

In comparison with our estimates of the island fractions, the two migration simulations in \citet{chi02}  
predict values of $f_\mathrm{S}$ = 0.16 and 0.13 and values of $f_\mathrm{L}$ = 0.48  and 0.23 for their slow migration (Ib) and fast migration (IIb) models, respectively.
Both of these results are rejected at the 99\% confidence level by the OSSOS population estimates.
A prediction of $f_\mathrm{S} = 0.42$ and $f_\mathrm{L} = 0.34$ for the low-inclination ($i < 5\arcdeg$) twotinos from Figure~9 of \citet{li14} is in agreement with our results. 
\citet{pike17b} provide an estimate of $f_\mathrm{S} = 0.37$$\pm0.07$ and $f_\mathrm{L} = 0.69$$\pm0.13$ (the uncertainties were estimated 
assuming poisson errors given the number of particles studied)
from a detailed Nice model simulation based on \citet{bra13}, which is rejected at more than 95\% confidence level by our analysis. 
\citet{law18c} included dynamical classifications of the four Neptune migration simulations in \citet{kai16} which used different migration timescales and either grainy or slow migration. This analysis provides a measure of $f_\mathrm{S}$/$f_\mathrm{L}$ for the twotinos in these four simulations: Grainy Slow = 0.42$\pm0.16$/0.36$\pm0.19$, Grainy Fast = 0.31$\pm0.05$/0.54$\pm0.08$, Smooth Slow = 0.13$\pm0.08$/0.52$\pm0.19$, Smooth Fast = 0.29$\pm0.05$/0.56$\pm0.09$ (again, uncertainties estimated assuming Poisson errors). 
Note that the slow migration simulations from \citet{law18c} have fewer particles in resonances, so the island fraction numbers are more uncertain than for the fast migration simulations.
Of these four studied cases, `Grainy Slow' migration provides the best match with our study, and the Smooth Slow migration simulation is rejected at $>$99\% confidence level.

As discussed in \citet{pike17b}, the specifics of Neptune’s migration as well as the test particle initial conditions have a significant effect on the relative populations of the twotino asymmetric islands. 
The dynamical processes to create the enhancement of libration island are still not comprehensively 
studied, and there is probably considerable uncertainty/variation in the model results themselves.
The range of $f_\mathrm{S}$ and $f_\mathrm{L}$ allowed by OSSOS will provide important constraints for future comprehensive numerical simulations of twotino production.

Outstanding issues that will hopefully be addressed in future Kuiper Belt surveys are:
(1) is there any difference in the population of the leading and trailing islands?  Our results
given no significant indication that there are (and a sample of hundreds of twotinos from a survey with very well
understood biases will be needed to convincingly demonstrate it), but if shown to exist this 
property might strongly constrain the style of planetary migration.
(2) Why is today's symmetric fraction so high?  There is no {\it a priori} reason to have thought that
the population is essentially equally split between symmetric, trailing, and leading librators,
but our results indicate this is roughly true.  Smooth slow migrations tend to produce a twotino
population poor in symmetric librators, so this may be a clue. 
Resonant sticking (see below) might serve to provide a steady state input of temporary resonators
that are more likely to be symmetric librators, but they would typically
have inclinations typical of scattering objects (15$^\circ$) than the cold distribution, limiting the allowed contribution of temporary twotinos.

\section{Discussion} \label{sec:dis}

The OSSOS survey's block right ascensions (RAs) was chosen to optimize the detection of resonant objects, especially twotinos. The RA of the C/D block is near the center of the leading island, and the  S, T, P and E blocks lean lightly against the lower boundaries of the leading and trailing islands (see Figure~\ref{fig:ossos_top_view}).
This survey design is very useful for probing the detailed structure of the twotino libration islands. Many OSSOS twotinos with $\phit\sim70\arcdeg$ and low $\dphit$ ($<10\arcdeg$) were discovered in the C/D block, which has excellent placement with deep discovery observations (longer exposure time), and the D block also benefited from better seeing conditions than the other blocks (see Figure~\ref{fig:fig_tracked_elibc}). 
These discoveries significantly constrain the distribution in phase space occupied by the current twotino population, especially their ranges of $\phit$ and $\dphit$.
The previous lack of low libration amplitude twotinos was thus a selection effect.
In the C/D block, OSSOS discovered six twotinos with $\dphit~<~10\arcdeg$, the lowest being $\dphit = 3.9\arcdeg$. However, the simulations of \citet{chi02} and \citet{pike17b} predict a much smaller population
of objects at such small libration amplitudes than that estimated from the OSSOS detections. The OSSOS results thus indicate that the initial conditions or migration parameters used in these migration simulations will need to be adjusted to produce more low libration amplitude twotinos.

OSSOS also revealed interesting regions of resonant parameter space that has not been sufficiently probed by observations.
o3o33 is a very peculiar object in the OSSOS sample. It has very low eccentricity (0.0825) and a unique intermediate libration amplitude of $113\arcdeg$. 
When we integrate the orbital evolution of o3o33, this object switches its libration center many times between the symmetric and asymmetric islands over the course of 10~Myr.
We note that 2001~UP$_{18}$, discovered by DES \citep{ell05}, also has a small eccentricity (0.0815) and is in the same process of switching islands. 
This behavior appears more common in this small low-$e$ portion of the twotino phase space.
Note that the low eccentricity also implies a larger observational bias against detecting such objects since the perihelion distance is larger.
The stability and the net flux between the symmetric and asymmetric islands are not clear, but they must have reached a steady state at the current time, given how rapidly the switching occurs.

The results of our globally statistical analysis are somewhat limited by having only \nt~OSSOS twotinos available. While this larger sample improves the uncertainties in the twotino population estimates by more than a factor of two compared to previous studies, the small number of detections still dominates the population uncertainties given in Section~\ref{sec:pop}. Nevertheless, our results allow an improved comparison of the OSSOS twotino population estimates to the plutino population estimate from \citet{vol16}. 
The estimated number of $10000^{+3600}_{-3000}$ plutinos brighter than $H_r < 8.66$ is somewhat more
than double the number of twotinos with the same $H_r$ limit, confirming the result of 
\citet{gla12}.
This low current ratio of the twotino to plutino populations must reflect a combination of their primordial population ratio (which depends on both the migration process and the primoridal TNO distribution at
the start of large-scale planetary migration) and their differential dynamical loss rate over the 
age of the Solar System.
\citet{hah05} demonstrate that the capture efficiency of the 2:1 resonance in Neptune’s migration (initial $\langle e \rangle= 0.001$, $\tau = 10^{7}$~yr) is significantly higher than that of the 3:2 resonance. 
\citet{tis09} present a simulation of this erosion process for both resonances, showing how the ancient ratio of the twotino and plutino populations can be related to the present ratio; 
the erosion rate of twotinos is about two times larger than that of the plutinos over 4~Gyr. 
Future numerical studies, incorporating the density/space distribution of the planetesimal disk, migration processes, erosion rate, and the estimates of the current plutino and twotino populations, would provide a cosmogonic picture to understand the history of the outer Solar System.

The inclination distribution in this study is consistent with the estimates of \citet{gla12} and \citet{vol16}; all studies prefer a colder inclination distribution for the twotinos than that of either the 3:2 or 5:2 resonances. 
This feature is very different from the slow migration simulations of \citet{nes15b}, which indicates a similar hotter inclination distribution for both the 3:2 and 2:1 resonances.
The fact that OSSOS detected very few high $i > 10\arcdeg$ twotinos, despite considerable 
sensitivity to them, indicates that the twotino population is intrinsically cold.
It is worth noting that \citet{tis09} identified inclination dependence in for long-term twotino stability,  which would mean that a sin($i$)-Gaussian distribution would not the best representation of the twotino inclination distribution even if the initially-trapped population was (see the bootstrapped AD test and the cumulative distribution in Figure~\ref{fig:fig_inc_prob} and \ref{fig:fig_cum_dieh}, which support this idea).
Although the inclination-dependent stability of twotinos in 1~Gyr timescale could contribute to a relatively cold inclination distribution \citep{tis09}, this effect is not noticeable in the results of \citet{nes15b}.
Further studies about how today's twotinos ($a\sim$~47~au) could be significantly dynamically colder in inclination than the two other largest resonance populations (3:2 $a\sim39$~au, 5:2 $a\sim55$~au) 
are needed, especially given that the twotino resonance is located between them.

Although we find that the existing detections do rule out many parametrizations of the intrinsic resonance distributions, there is still freedom related to how the libration centers and amplitudes can be modeled/adjusted.
We are convinced, however, that this will not greatly alter the population estimates, considering the comparable population estimations with a relatively simple twotino model in \citet{gla12} and \citet{vol16}.
More cosmogonic numerical studies of early Solar System dynamics are required in order to determine how much variation there is in the current
epoch's distribution of $\langle\phit\rangle$ and $\dphit$, and how these distributions depend on migration rates, TNO initial conditions, and long-term dynamical erosion.  
With only the existing numerical studies, we are unable to precisely determine if variations in the early Solar 
System (for example, the rate and extent of planetary migration, or the $e$ and $i$ distribution of small TNOs at the start of planet migration) result in different distributions today.

Future studies of the 2:1 resonance are probably not warranted until at least a factor of two more objects are discovered. This will likely happen once the Large Synoptic Survey Telescope (LSST) \citep{ive18} begins operations, which is the only project in the foreseeable horizon with the scope to make a major leap forward in the number of detections.
Although OSSOS was a deeper survey than LSST, the roughly two orders of magnitude more sky area to be covered by LSST will more than compensate for this, and the survey should detect several hundred twotinos. 
It will be critically important for the LSST survey to ensure that the completeness of the TNO detections is known, as OSSOS makes it clear that the number of detections is a {\it very} strong function of depth at
various sky locations ({\it e.g.,} Figure~\ref{fig:ossos_top_view} shows that the observations of the deepest OSSOS block, labeled C/D, were very good at finding twotinos, especially leading asymmetric librators). The LSST project would thus need to invest significant effort into understanding if there is any longitudinal sensitivity to its detection probability due to seasonal weather trends and background stellar density.

\section{Summary} \label{sec:sum}

The empirical model developed for this work acceptably reproduces the distribution of OSSOS twotino detections in all of the parameters studied. 
This match allows us to confidently derive population estimates for the 2:1 populations in each libration island.
Note that the AD statistic provides the non-rejectable range of a parameter, rather than assessing the best value of a parameter.
We summarize our main findings as follows:

\begin{enumerate}
\item
Although the model presented in \citet{vol16} is rejected after the inclusion of the full OSSOS twotino sample, it proves we can probe the dynamical structure within the islands.
Our model parameters and population estimates are fully validated with the OSSOS survey simulator to match with the detected OSSOS twotino sample.
These are the first results from a fully characterized survey with a large enough sample population to investigate the full twotino parameter space and the differences in the distributions between the different libration islands.

\item
Our analysis strongly supports previous claims \citep{gla12, vol16} that the width $\sigma_\mathrm{i}$ of the twotino population inclination distribution is significantly colder than other resonances. \citet{vol16} shows a best estimate of $\sigma_\mathrm{i} = 12\arcdeg$ for the 3:2 resonance width, while our estimate for the twotino population is $\sigma_\mathrm{i} = 6\arcdeg$. These distributions are different at more than 95\% confidence level.

\item
The combination of the empirical model and the OSSOS survey simulator provides an excellent match to the observed twotino population. However, more detections, especially of fainter and more distant twotinos, would be very helpful to further constrain the parameter space for comparison to migration models.

\item
Our estimates of the overall distributions of libration centers and amplitudes
are in good agreement with the simulations of \citet{chi02} and \citet{li14}, with the exception of the low libration amplitude objects. 
Our debaised results indicate that the current populations of the symmetric, leading, and trailing
islands are all equal given current uncertainties.
Our estimates of the libration island fractions appear discrepant with some models of outward Neptune migration, especially slow and smooth ones.

\end{enumerate}

\acknowledgments

The authors acknowledge the staff of the Canada–France–Hawaii Telescope for their amazing effort in this project. 
We are grateful to Jian Li for kindly providing the data from his study of the twotino inclination distribution. Y.-T.C. gratefully acknowledges support from NRC-Canada and UBC, enabling a scholarly visit in the early phase of this study.
Y.-T.C. also acknowledge Renu Malhotra for helpful conversations.
KV and RMC acknowledge support from NASA grant NNX15AH59G. KV acknowledges additional support from NASA grant NNX14AG93G and NSF grant AST-1824869.
B.G. acknowledges NSERC Canada for research support.
M.T.B. acknowledges support from UK Science and Technology Facilities Council grant ST/P0003094/1.
The Center for Exoplanets and Habitable Worlds is supported by the Pennsylvania State University, the Eberly College of Science, and the Pennsylvania Space Grant Consortium.
S.M.L. gratefully acknowledges support from the NRC-Canada Plaskett Fellowship.
R.I.D. is supported in part by Alfred P. Sloan Foundation's Research Fellowship.
Based on observations obtained with MegaPrime/MegaCam, a joint project of CFHT and CEA/DAPNIA, at the Canada-France-Hawaii Telescope (CFHT) which is operated by the National Research Council (NRC) of Canada, the Institut National des Science de l'Univers of the Centre National de la Recherche Scientifique (CNRS) of France, and the University of Hawaii, with this project receiving additional access due to contributions from the Institute of Astronomy and Astrophysics, Academia Sinica, National Tsing Hua University, and Ministry of Science and Technology, Taiwan.

\facilities{CFHT (MegaCam)}
\software{orbfit \citep{ber00}, MERCURY \citep{cha99} and OSSOS Survey Simulator \citep{ban16, law18}}

\bibliographystyle{aasjournal}
\bibliography{sample61}

\end{document}